\documentclass[dvipsnames,preprint,12pt,sort&compress]{elsarticle}

\usepackage{amssymb}

\usepackage{amsthm}
\usepackage{amstext}
\usepackage{amsmath}
\usepackage{amsfonts}
\usepackage{listings}
\usepackage{color}
\usepackage{tikz}
\usepackage{graphicx}
\usepackage{caption}
\usepackage{xspace}

\newtheorem{theorem}{Theorem}

\newcounter{bla}

\journal{Computer Physics Communications}

\def\d{{\rm d}}

\lstdefinestyle{Common}
{
    basicstyle=\footnotesize \ttfamily\null,
    numbers=none,
    numbersep=1em,
    frame=single,
    framesep=\fboxsep,
    framerule=\fboxrule,
    xleftmargin=\dimexpr\fboxsep+\fboxrule,
    xrightmargin=\dimexpr\fboxsep+\fboxrule,
    breaklines=true,
    breakindent=0pt,
    tabsize=5,
    columns=flexible,
    showstringspaces=false,
    captionpos=b,
    abovecaptionskip=0.2\smallskipamount,
}

\lstdefinestyle{Mathematica}
{
    style=Common,
    language={Mathematica},
    alsolanguage={[LaTeX]TeX},
    keywordstyle=\color{Black},
    identifierstyle=\color{RoyalBlue},
    numberstyle=\color{Green},
    commentstyle=\color{gray},
    stringstyle=\color{black},
    morekeywords=
    {
	MultivariateResidues,
	MultivariateResidue,
	MultiResInputChecks,
	MultiResInternalChecks,
	FindMinimumDelta,
	MultiResUseSingular,
	MultiResSingularPath,
    GlobalResidueTheoremCPn,
    },
    morecomment=[l]{Out:}
}

\lstdefinestyle{CodeSample}
{
    style=Mathematica,
}

\newcommand{\Math}[1]
{\lstinline[style=Mathematica,breaklines=false,basicstyle=\small \ttfamily\null]~#1~}

\lstnewenvironment{CodeSample}[1][]
{\vspace{-3mm}\lstset{style=CodeSample,#1}}
{}

\begin{document}

\begin{frontmatter}



\title{\textsc{MultivariateResidues:} a Mathematica package \\ for computing multivariate residues}


\author[a,b]{Kasper J. Larsen\corref{author}}
\author[c,d,e]{Robbert Rietkerk}\ead{Robbert.Rietkerk@kit.edu}

\cortext[author] {Corresponding author. \textit{E-mail address:} \texttt{Kasper.Larsen@soton.ac.uk}}
\address[a]{Institute for Theoretical Physics, ETH Z{\"u}rich, 8093 Z{\"u}rich, Switzerland}
\address[b]{School of Physics and Astronomy, University of Southampton, \\ Highfield, Southampton, SO17 1BJ, United Kingdom}

\address[c]{Nikhef, Theory Group, Science Park 105, 1098 XG Amsterdam, The Netherlands}
\address[d]{Institute for Theoretical Physics, University of Amsterdam, Science Park 904, \\ 1098 XH Amsterdam, The Netherlands}
\address[e]{Institute for Theoretical Particle Physics, KIT, Wolfgang-Gaede-Strasse 1, \\ 76128 Karlsruhe, Germany}

\begin{abstract}
Multivariate residues appear in many different contexts
in theoretical physics and algebraic geometry. In theoretical physics,
they for example give the proper definition of generalized-unitarity cuts,
and they play a central role in the Grassmannian formulation of the S-matrix
by Arkani-Hamed et al. In realistic cases their evaluation can be non-trivial.
In this paper we provide a Mathematica package for efficient evaluation
of multidimensional residues based on methods from computational algebraic geometry.
The package moreover contains an implementation of the global residue theorem,
which produces relations between residues at finite locations and residues at infinity.
\end{abstract}


\end{frontmatter}

\raisebox{155mm}[0mm][0mm]{\hspace{-6mm}
Nikhef-2016-058, TTP16-062
}\vspace{-2mm}



\noindent {\bf PROGRAM SUMMARY} \\

\begin{small}
\noindent
{\em Program Title:} \textsc{MultivariateResidues}                    \\
{\em Licensing provisions:} GNU General Public License (GPL)          \\
{\em Programming language:} Wolfram Mathematica version 7.0 or higher \\
{\em Nature of problem:} Evaluation of multivariate complex residues  \\
{\em Solution method:} Mathematica implementation
\end{small}


\section{Introduction}\label{sec:introduction}

Multivariate residues appear in many different contexts
in theoretical physics and algebraic geometry. In theoretical physics,
they for example give the proper definition of generalized-unitarity cuts%
~\cite{Cachazo:2008dx,Cachazo:2008vp,Kosower:2011ty,CaronHuot:2012ab,Sogaard:2013fpa,Johansson:2015ava},
and they play a central role in the Grassmannian formulation of the S-matrix
by Arkani-Hamed et al.~%
\cite{ArkaniHamed:2009dn,ArkaniHamed:2009dg,ArkaniHamed:2010gh,ArkaniHamed:2012nw,Arkani-Hamed:2013kca,Arkani-Hamed:2014bca,Arkani-Hamed:2014dca,Bern:2015ple,Herrmann:2016qea}. %
A recent paper \cite{Chen:2017bug} uses multivariate residues to construct Bern-Carrasco-Johansson numerators \cite{Bern:2008qj,Bern:2010ue} for gauge theory loop integrands.
In algebraic geometry, multivariate residues play an important role
in elimination theory in the context of solving systems of
multivariate polynomial equations \cite{CattaniDickenstein}.

In practice, the evaluation of multivariate residues can be non-trivial.
Nevertheless, implementations of their evaluation have not been made
publically available. In this paper we provide the Mathematica package
\textsc{MultivariateResidues} for efficient evaluation of multivariate residues based
on methods from computational algebraic geometry.
Related work has recently appeared in the package \textsc{Rings} \cite{Poslavsky:2017sne}
which provides a library for computing factorization, GCDs etc.~of multivariate polynomials
over arbitrary coefficient rings.

This paper is organized as follows. In section~\ref{sec:General_theory}
we give the definition of the multivariate residue along with some of its basic
properties, in particular the transformation formula. We explain an algorithm
for how the latter can be utilized to compute multivariate residues in general. In
section~\ref{sec:quotient_ring_duality} we explain an alternative approach which
makes use of powerful methods from modern commutative algebra. Both of these methods
are implemented in \textsc{MultivariateResidues}. In section~\ref{sec:example} we
apply the formalism of section~\ref{sec:quotient_ring_duality} to a specific example
to illustrate how residues are computed in practice.
In section~\ref{sec:global_residue_thm} we discuss the global residue theorem.
In section~\ref{sec:applications} we discuss the application of multivariate residues
to the calculation of generalized-unitarity cuts in the context of computations
of scattering amplitudes in perturbative quantum field theory. Section~\ref{sec:manual} provides a manual
for \textsc{MultivariateResidues} along with benchmarks of the performance,
comparisons between the various options and tips for the user to improve performance.
In section~\ref{sec:Conclusions} we give our conclusions.
\ref{sec:topology_of_multivariate_residues} provides a topological explanation
of why multivariate residues, in contrast to the univariate case, are
not uniquely determined by the location of a pole, but have some dependence on the
integration cycle.

\section{General theory}\label{sec:General_theory}

In this section we give the definition of multivariate complex
residues and discuss the transformation formula and how this may
be utilized to compute residues in practice.

Our setup is as follows. Let $f(z) = \big(f_1(z), \ldots, f_n(z) \big) : \mathbb{C}^n \to \mathbb{C}^n$
and $h : \mathbb{C}^n \to \mathbb{C}$ be holomorphic functions, and
consider the meromorphic $n$-form,%
\footnote{If one adds a boundary at infinity as needed to apply
global residue theorems, we can define the functions on $\mathbb{CP}^n$
rather than $\mathbb{C}^n$.}
\begin{align}
\omega = \frac{h(z) \hspace{0.4mm} \d z_1\wedge\cdots\wedge \d z_n}{f_1(z)\cdots f_n(z)}\,.
\label{eq:diff_form_generic}
\end{align}
The case where the form has $m$ denominator factors with $m>n$ can be treated as a special
case of the above by grouping the $m$ factors into precisely $n$ factors.
We will elaborate on the ambiguity of this process and its underlying
topological explanation further below. Likewise, the case of
$m < n$ denominator factors will be discussed below.

In the multivariate setting, we define a pole as a point $p \in \mathbb{C}^n$
where $f$ has an isolated zero---that is, $f(p) = 0$ and $f^{-1}(0)\cap U = \{p\}$ for
a sufficiently small neighborhood $U$ of $p$.
We are interested in computing the residue of $\omega$ at its poles.
The multivariate residue is defined as a multidimensional generalization
of a contour integral: an integral taken over a product of $n$ circles,
that is an $n$-torus,
\begin{align}
\mathop{\mathrm{Res}}_{\{f_1,\dots,f_n\}, \hspace{0.6mm} p}(\omega) =
\frac{1}{(2\pi i)^n}
\oint_{\Gamma_\epsilon}
\frac{h(z) \hspace{0.6mm} \d z_1\wedge\cdots\wedge \d z_n}{f_1(z)\cdots f_n(z)}\,,
\label{eq:residue_def}
\end{align}
where $\Gamma_\epsilon = \{z\in \mathbb{C}^n:|f_i(z)| = \epsilon_i\}$ and the $\epsilon_i$
have infinitesimal positive values. Furthermore, the integration cycle is oriented by
the condition,
\begin{equation}
\d (\mathrm{arg} \hspace{0.6mm} f_1) \wedge \cdots \wedge \d (\mathrm{arg} \hspace{0.6mm} f_n) \geq 0 \,.
\label{eq:orientation_of_cycle}
\end{equation}
We note that the definition of the integration cycle differs from the univariate case:
rather than being defined directly in terms of the variables $z$, $\Gamma_\epsilon$ is
defined in terms of the denominator factors $f_i(z)$.

The Jacobian determinant evaluated at the pole
\begin{align}
J(p)\equiv \det_{i,j}\left(  \frac{\partial f_i}{\partial z_j}  \right)\bigg|_{z = p}\,,
\label{eq:Jacobian_determinant}
\end{align}
plays an important role, since if $J(p) \neq 0$, we can evaluate the
residue directly by the coordinate transformation $w = f(z)$,
\begin{align}
\mathop{\mathrm{Res}}_{\langle f_1,\dots,f_n \rangle,p}(\omega)
= \frac{1}{(2\pi i)^n}
\oint_{|w_i| \leq \epsilon_i} \hspace{-1.5mm}
\frac{h\big(f^{-1}(w)\big) \hspace{0.6mm} \d w_1\wedge\cdots\wedge \d w_n}{J(p) \hspace{0.6mm} w_1 \cdots w_n}
= \frac{h(p)}{J(p)} \,.
\label{eq:residue_from_Jacobian}
\end{align}
In this case, the residue is termed nondegenerate.

In general, however, a residue may be degenerate, such as is the
case for higher-order poles. In this situation, the above
coordinate transformation does not suffice to compute it.
A central and completely general property of residues
is the transformation formula (cf.~section 5.1 of ref.~\cite{GriffithsHarris}).
As we will shortly see, this property can be utilized to
compute any residue, degenerate or nondegenerate.

\begin{theorem}
(Transformation formula). Let $I = \langle f_1 (z), \ldots, f_n (z) \rangle$
be a zero-dimensional ideal%
\footnote{The ideal $I$ is said to be \emph{zero-dimensional}
if and only if the solution to the equation system $f_1 (z) = \cdots = f_n (z) = 0$
consists of a finite number of points $z \in \mathbb{CP}^n$.} generated
by a finite set of holomorphic functions $f_i (z) : \mathbb{CP}^n \to \mathbb{C}$
with $f_i (p) = 0$. Furthermore, let $J = \langle g_1 (z), \ldots, g_n (z) \rangle$
be a zero-dimensional ideal such that $J \subseteq I$; that is, whose generators
are related to those of $I$ by $g_i (z) = \sum_{j=1}^n a_{ij} (z) f_j(z)$
with the $a_{ij} (z)$ being holomorphic functions. Letting $A(z) = \big(a_{ij} (z) \big)_{i,j=1,\ldots,n}$
denote the transformation matrix, the residue at $p$ satisfies,
\end{theorem}

\vspace{-5mm}

\begin{equation}
\mathop{\mathrm{Res}}_{\langle f_1, \ldots, f_n \rangle, \hspace{0.3mm} p} \hspace{-0.2mm}\left(
\frac{h(z) \hspace{0.4mm} \d z_1 \wedge \cdots \wedge \d z_n}{f_1 (z) \cdots f_n (z)} \right)
\hspace{1mm}=\hspace{1mm} \mathop{\mathrm{Res}}_{\langle g_1, \ldots, g_n \rangle, \hspace{0.3mm} p}
 \hspace{-0.2mm}\left(
\frac{h(z) \hspace{0.1mm} \det A(z) \hspace{0.4mm} \d z_1 \wedge \cdots \wedge \d z_n}
{g_1 (z) \cdots g_n (z)} \right) \,. \label{eq:transformation_law}
\end{equation}

In cases where the form $\omega$ has fewer denominator factors than variables,
the notion of residue defined in eq.~\eqref{eq:residue_def} does not apply,
and this case is therefore outside the scope of this paper. Nevertheless,
we mention that a notion of residue which does apply in this situation is
that of \emph{residual forms}. To illustrate the idea, let us consider
the following example, taken from section 7.2 of ref.~\cite{ArkaniHamed:2009dn},
\begin{equation}
\omega = \frac{\d z_1 \wedge \d z_2 \wedge \d z_3}{z_1 (z_1 + z_2 z_3)} \,.
\end{equation}
As $\omega$ has three variables, but only two denominator factors, the
residue in eq.~\eqref{eq:residue_def} is not well-defined. However, we observe
that we can define the 2-form
\begin{equation}
\widetilde{\omega} \hspace{0.8mm}=\hspace{0.8mm} \mathop{\mathrm{Res}}_{z_1=0} \hspace{0.4mm} \omega
\hspace{0.8mm}=\hspace{0.8mm} \frac{\d z_2 \wedge \d z_3}{z_2 z_3} \,.
\end{equation}
This form has two variables and two denominator factors, and hence the notion
of residue in eq.~\eqref{eq:residue_def} applies to $\widetilde{\omega}$.

\subsection{Computation of residues via the transformation formula}%
\label{sec:residues_from_transformation_formula}

To apply the transformation formula \eqref{eq:transformation_law} to
the computation of residues, we must first find a useful transformation
of the set of ideal generators. Here we restrict attention to the case
where the generators $f_i(z)$ are polynomials and follow the approach explained
in section~1.5.4 of ref.~\cite{CattaniDickenstein}.
The idea is to choose the $g_i$ to be \emph{univariate}---that is,
$g_i (z_1, \ldots, z_n) = g_i(z_i)$. Then the residue can simply
be evaluated as a product of univariate residues.

A set of univariate polynomials $g_i$ can be obtained by generating
a Gr{\"o}bner basis
of $\{ f_1(z), \ldots, f_n(z) \}$ with lexicographic monomial order.
Specifying the variable ordering $z_{i+1} \succ z_{i+2} \succ \cdots \succ z_n \succ z_1
\succ z_2 \cdots \succ z_i$ will produce a Gr{\"o}bner basis
whose first element is a polynomial which depends only on $z_i$.
We let $g_i (z_i)$ denote this polynomial.
Now, by considering all $n$ cyclic permutations of the variable ordering
$z_1 \succ z_2 \succ \cdots \succ z_n$ in this way we generate a set of $n$
univariate polynomials $\{ g_1(z_1), \ldots, g_n(z_n) \}$.

To illustrate the above method, we consider as an example
the following differential form,
\begin{align}
\omega = \frac{z_1 \hspace{0.6mm} \d z_1\wedge \d z_2}{z_2 (a_1 z_1 + a_2 z_2)(b_1 z_1 + b_2 z_2)} \,,
\label{eq:degenerate_diff_form}
\end{align}
which at the same time will serve to explain how to compute residues in
cases with more distinct denominator factors than variables.
As eq.~\eqref{eq:degenerate_diff_form} depends on two variables and
has three distinct denominator factors, we must consider all possible
ways of partitioning the denominator into two factors. Denoting
the denominator factors of eq.~(\ref{eq:degenerate_diff_form}) as follows,
\begin{equation}
\begin{aligned}
\varphi_1 (z_1,z_2)  &=  z_2 \\
\varphi_2 (z_1,z_2)  &=  a_1 z_1 + a_2 z_2 \\
\varphi_3 (z_1,z_2)  &=  b_1 z_1 + b_2 z_2 \,,
\end{aligned}
\label{eq:denominator_factors_example}
\end{equation}
we observe that this can be done in three distinct ways, namely
\begin{equation}
\{ \varphi_1, \varphi_2 \varphi_3 \}\,, \hspace{5mm} \{ \varphi_2, \varphi_3 \varphi_1 \}
\hspace{5mm} \mathrm{and} \hspace{5mm} \{ \varphi_3, \varphi_1 \varphi_2 \} \,.
\label{eq:denominator_partitionings_example}
\end{equation}
We are interested in computing the residues of $\omega$ at the pole
$p = (0,0)$ corresponding to each of these partitionings. We note
that all of these residues are degenerate.

Let us evaluate the residue for the denominator partitioning $\{ \varphi_1, \varphi_2 \varphi_3 \}$.
The lexicographically-ordered Gr{\"o}bner basis of $\{ \varphi_1, \varphi_2 \varphi_3 \}$ in the variable ordering
$z_2 \succ z_1$ is $\{ a_1 b_1 z_1^2, z_2 \}$; in the variable
ordering $z_1 \succ z_2$ it is $\{ z_2, a_1 b_1 z_1^2 \}$. Choosing the first
element of each Gr{\"o}bner basis we obtain,
\begin{align}
g_1(z_1,z_2)  &=  a_1 b_1 z_1^2 \\
g_2(z_1,z_2)  &=  z_2 \,.
\end{align}
We can obtain the transformation matrix $A$ as a byproduct of finding the Gr{\"o}bner
basis (or using the approach implemented in ref.~\cite{Lichtblau}). In the simple case
considered here, ordinary multivariate polynomial division produces the same
result,
\begin{equation}
A \hspace{0.7mm}=\hspace{0.7mm} \begin{pmatrix} -(a_1 b_2 + a_2 b_1) z_1 - a_2 b_2 z_2 & \hspace{1.5mm} 1 \\[1mm]
1 & \hspace{1.5mm} 0 \end{pmatrix}\,,
\end{equation}
that relates the two sets of ideal generators,
\begin{equation}
A \cdot \hspace{0.4mm} \begin{pmatrix} \varphi_1(z_1, z_2) \\[1mm] \varphi_2(z_1, z_2) \varphi_3(z_1, z_2) \end{pmatrix}
\hspace{1mm}=\hspace{1mm} \begin{pmatrix} g_1(z_1, z_2) \\[1mm] g_2(z_1, z_2) \end{pmatrix} \,.
\end{equation}
From the transformation law (\ref{eq:transformation_law}) we then find that the residue of $\omega$
at $p = (0,0)$ with respect to the ideal generators $\{ \varphi_1, \varphi_2 \varphi_3 \}$ is
\begin{equation}
\mathop{\mathrm{Res}}_{\{ \varphi_1, \varphi_2 \varphi_3 \}, \hspace{0.8mm} p} \omega
\hspace{1.5mm}=\hspace{1.5mm} \mathop{\mathrm{Res}}_p \frac{z_1 \hspace{0.5mm}
\det A \hspace{1.6mm} \d z_1 \wedge \d z_2}{g_1(z_1, z_2) \hspace{0.5mm} g_2 (z_1, z_2)}
\hspace{1.5mm}=\hspace{1.5mm} - \mathop{\mathrm{Res}}_p \frac{\d z_1 \wedge \d z_2}
{a_1 b_1 z_1 z_2} \,.
\label{eq:residue_omega_{phi1_phi2phi3}}
\end{equation}
As desired, the denominator on the right-hand side of eq.~(\ref{eq:residue_omega_{phi1_phi2phi3}})
is a product of univariate polynomials. Hence the residue can
be computed as a product of univariate residues and yields,
\begin{align}
R_1 \hspace{0.8mm}\equiv\hspace{0.8mm} \mathop{\mathrm{Res}}_{\{ \varphi_1, \varphi_2 \varphi_3 \}, \hspace{0.5mm} p} \omega
\hspace{1mm}&=\hspace{1mm} -\frac{1}{a_1 b_1}  \label{eq:R_1_def} \\
R_2 \hspace{0.8mm}\equiv\hspace{0.8mm} \mathop{\mathrm{Res}}_{\{ \varphi_2, \varphi_3 \varphi_1 \}, \hspace{0.5mm} p} \omega
\hspace{1mm}&=\hspace{1mm} -\frac{a_2}{a_1 (a_1 b_2 - a_2 b_1)} \label{eq:R_2_def} \\
R_3 \hspace{0.8mm}\equiv\hspace{0.8mm} \mathop{\mathrm{Res}}_{\{ \varphi_3, \varphi_1 \varphi_2 \}, \hspace{0.5mm} p} \omega
\hspace{1mm}&=\hspace{1mm} \frac{b_2}{b_1 (a_1 b_2 - a_2 b_1)} \,, \label{eq:R_3_def}
\end{align}
where the residues for the two other denominator partitionings $\{ \varphi_2, \varphi_3 \varphi_1\}$
and $\{ \varphi_3, \varphi_1 \varphi_2\}$ were computed analogously.
We remark that in general one must keep in mind that the residue
is antisymmetric under interchanges of the denominator factors of $\omega$.
This follows from the dependence of the residue on the orientation of
the integration cycle, cf.~eq.~\eqref{eq:orientation_of_cycle}.

We observe that only two out of the three residues $R_1, R_2, R_3$
in eqs.~(\ref{eq:R_1_def})--(\ref{eq:R_3_def}) are independent,
as the residues satisfy the identity,
\begin{equation}
R_1 + R_2 + R_3 = 0 \,.
\label{eq:residue_relation}
\end{equation}
Identities of this kind are common for multivariate residues.

In \ref{sec:topology_of_multivariate_residues} we give a topological explanation
of why the multivariate residues in eqs.~\eqref{eq:R_1_def}--\eqref{eq:R_3_def}
are not uniquely determined by the pole $p$, but rather also depend on the
choice of partitionings in eq.~\eqref{eq:denominator_partitionings_example}.

\section{Evaluation of residues by use of dual structure of quotient ring}\label{sec:quotient_ring_duality}

The evaluation of residues by use of the transformation formula explained in
section~\ref{sec:residues_from_transformation_formula}
is completely general. However, in realistic cases the computation of
the transformation matrix $A$ can be intensive, and as a result this method
is not optimal in all situations.

In this section we explain a more efficient method for residue computations,
which we have implemented in \textsc{MultivariateResidues}. Our setup is as follows.
As in section~\ref{sec:General_theory}, we restrict ourselves to the case
where the denominator factors of $\omega$ in eq.~\eqref{eq:diff_form_generic}
are polynomials. We denote these polynomials by $P_1(z), \ldots, P_n(z)$ and
assume that the ideal $I=\langle P_1(z), \ldots, P_n(z) \rangle$ is zero-dimensional;
i.e., that the associated variety
$V = \{ z \in \mathbb{C}^n : \hspace{0.7mm} P_1(z) = \cdots = P_n(z) = 0 \}$
consists of a finite number of points,
\begin{equation}
V = \{ p_1, \ldots, p_m \} \,.
\end{equation}

This method exploits
that the residue map defines a non-degenerate inner product $\langle \cdot, \cdot \rangle$
on the quotient ring
\begin{equation}
Q \equiv \mathbb{C}[z_1, \ldots, z_n]/I
\end{equation}
of the ring $\mathbb{C}[z_1, \ldots, z_n]$ of all polynomials in the
variables $z_1, \ldots, z_n$ with coefficients in $\mathbb{C}$ modulo
the ideal $I$. As $I$ is zero-dimensional, $Q$ has a finite dimension
(cf.~section 2.2 of ref.~\cite{UsingAlgebraicGeometry}) which we denote by $D$.

By decomposing the numerator of $\omega$ in a canonical
(linear) basis of the quotient ring, and the constant 1
in the dual basis wrt.~$\langle \cdot, \cdot \rangle$,
the \emph{global residue}
\begin{equation}
\mathop{\mathrm{Res}}_I \hspace{0.5mm} \omega
\hspace{1.2mm}=\hspace{1.8mm} \sum_{i=1}^m \hspace{0.6mm} \mathop{\mathrm{Res}}_{I, \hspace{0.6mm} p_i} \hspace{0.5mm} \omega
\label{eq:global_residue_def}
\end{equation}
can be computed as the dot product of the corresponding
coefficient vectors. The corresponding local residue at each pole $p_i$
can then be computed by multiplying the integrand by appropriate polynomials
which are unity in the vicinity of $p_i$ and vanish in the vicinity of
the remaining poles.

In the following we explain how the canonical and dual (linear) bases are computed
and how the above-mentioned polynomials are constructed.

\subsection{Computing the canonical basis of the quotient ring}

Our first aim is to determine a canonical (linear) basis of the quotient ring $Q$.
To this end we compute a Gr{\"o}bner basis $G$ of $I$ and consider the
ideal $\langle LT(I) \rangle$ generated by the leading term of each element of $G$.
The monomials in the complement of $\langle LT(I) \rangle$ then form a basis of $Q$.
(Cf.~proposition 1 of section 5.3 of ref.~\cite{IdealsVarietiesAlgorithms}.)

We can rephrase this statement as the following algorithm.

\begin{enumerate}
\item  Decide on a monomial order $\prec$ and compute a Gr{\"o}bner basis
$G = \{ g_1, \ldots, g_s\}$ of $I$ wrt. $\prec$.\\[1mm]

\item  Obtain the leading term of each Gr{\"o}bner basis element,
\begin{equation}
h_i = \mathrm{LT}(g_i) \,.
\end{equation}
\vspace{-5mm}

\item Extract the exponent vectors of the leading terms
\begin{equation}
h_i = z_1^{\alpha_{i,1}} \cdots z_n^{\alpha_{i,n}} \hspace{6mm} \longmapsto \hspace{6mm} (\alpha_{i,1}, \ldots, \alpha_{i,n}) \,.
\end{equation}
\vspace{-5mm}

\item The elements of
\begin{equation}
E \hspace{1mm}=\hspace{1mm} \mathbb{Z}_{\geq 0}^n \Big\backslash
\hspace{1.2mm} \bigcup_{i=1}^s \big(  (\alpha_{i,1}, \ldots, \alpha_{i,n}) + \mathbb{Z}_{\geq 0}^n  \big)
\label{eq:exponent_vectors_of_can_basis}
\end{equation}
then define exponent vectors of the canonical basis elements. That is,
\begin{equation}
\mathcal{C} = \big\{ z_1^{\beta_{1,1}} \hspace{-0.5mm} \cdots \hspace{-0.5mm} z_n^{\beta_{1,n}},
\hspace{2mm} \ldots, \hspace{2mm} z_1^{\beta_{D,1}} \hspace{-0.5mm} \cdots \hspace{-0.5mm}
z_n^{\beta_{D,n}} \big\} \hspace{5mm} \mathrm{where} \hspace{4mm}  (\beta_{j,1}, \ldots, \beta_{j,n}) \in E \,,
\end{equation}
is the desired canonical basis of the quotient ring $Q$ wrt. $\prec$.
\end{enumerate}

\subsection{Computing the dual basis of the quotient ring}

Our next aim is to determine the dual (wrt.~$\mathcal{C}$) basis of $Q$.
This basis can be extracted from the determinant of the Bezoutian matrix
of the polynomials $P_1 (z), \ldots, P_n (z)$.
Accordingly, we proceed with the following steps.

\begin{enumerate}
\item Compute the Bezoutian matrix of the polynomials $P_1 (z), \ldots, P_n (z)$,
\begin{equation}
\mathrm{Bez}_{ij} (z,y) \hspace{1mm}=\hspace{1mm} \frac{P_i (y_1, \ldots, y_{j-1}, z_j, \ldots, z_n) - P_i (y_1, \ldots, y_j, z_{j+1}, \ldots, z_n)}{z_j - y_j} \,.
\label{eq:Bezoutian_matrix_definition}
\end{equation}
The entries of the Bezoutian matrix are elements of the direct product $Q \otimes Q$.

\item Take the determinant of the Bezoutian matrix
\begin{equation}
\mathcal{B} (z, y) \hspace{0.5mm}\equiv\hspace{0.5mm} \det (\mathrm{Bez}) \,.
\label{eq:Bezoutian_determinant_definition}
\end{equation}

\item Compute the remainder of $\mathcal{B} (z, y)$ in $Q \otimes Q$.
This is carried out in practice by first performing polynomial division of $\mathcal{B} (z, y)$
wrt.~the Gr{\"o}bner basis
$G = \{ g_1(z), \ldots, g_s(z)\}$ where
the elements are taken as polynomials in $z_1, \ldots, z_n$, and then
performing polynomial division of the result wrt. $G$ whose elements are now taken as polynomials in $y_1, \ldots, y_n$,
\begin{align}
\mathcal{B} (z, y)   \hspace{1mm} \label{eq:pol_div_of_Bez_wrt_z}
&=\hspace{1mm} q_1 (z,y) g_1 (z) + \cdots + q_s (z,y) g_s (z) + \mathcal{B}_Q (z, y) \,, \\
\mathcal{B}_Q (z, y) \hspace{1mm}&=\hspace{1mm} \widehat{q}_1 (z,y) g_1 (y) + \cdots + \widehat{q}_s (z,y) g_s (y) + \mathcal{B}_{Q \otimes Q} (z, y) \,. \label{eq:pol_div_of_Bez_wrt_y}
\end{align}

\item Label the elements of the canonical basis as
$\mathcal{C} = \{ c_1 (z), \ldots, c_D (z)\}$ and decompose the
Bezoutian determinant as,
\begin{equation}
\mathcal{B}_{Q \otimes Q} (z, y) = c_1 (z) d_1 (y)
+ \cdots + c_D (z) d_D (y) \,.
\end{equation}
$\mathcal{B}_{Q \otimes Q}$ has a unique such decomposition, and the dual basis
of $Q$ (wrt. the canonical basis $\mathcal{C}$) can now be read off
(cf.~section 1.5.4 of ref.~\cite{CattaniDickenstein}),
\begin{equation}
\mathcal{D} = \{ d_1 (z), \ldots, d_D (z) \} \,,
\end{equation}
where the variables were relabeled into $z_1, \ldots, z_n$. We remark that the
elements $d_i (z)$ are in general polynomials
rather than monomials (in contrast to the canonical basis elements).
\end{enumerate}

\subsection{Constructing partition-of-unity polynomials}

One more ingredient is needed to compute residues at all the poles in
the variety $V = \{ p_1, \ldots, p_m \}$ associated with $I$, namely a set of
polynomials $e_1(z), \ldots, e_m(z)$
which are unity in the vicinity of a given pole and vanishing in the vicinity
of the remaining poles.

To this end, we construct a linear form $\ell (z) = a_1 z_1 + \cdots + a_n z_n$
(with $a_i \in \mathbb{C}$) such that $\ell(p_1), \ldots, \ell(p_m)$ are all distinct.
(In practice, this is done in \textsc{MultivariateResidues}\- by scanning over a set
of coefficient vectors $(a_1, \ldots, a_n)$ with integer entries.)

The following set of Lagrange polynomials
\begin{equation}
L_i (z) = \prod_{\substack{j=1, \\ j \neq i}}^m \frac{\ell(z - p_j)}{\ell(p_i - p_j)}
\label{eq:Lagrangian_polynomials_generic_formula}
\end{equation}
then have the desired property of ``projecting onto each pole'',
\begin{equation}
L_i (p_k) = \delta_{ik} \,.
\label{eq:partition-of-unity_property_1}
\end{equation}
However, this set of polynomials will not quite have the desired
property of defining a partition of unity,
\begin{equation}
\sum_{i=1}^m e_i = 1 \hspace{3mm} \mbox{(mod $I$) \hspace{4mm} and}
\hspace{6mm} e_i e_j = e_i \delta_{ij} \hspace{3mm} \mbox{(mod $I$)} \,.
\label{eq:partition-of-unity_property_2}
\end{equation}
Rather (cf.~lemma 2.3 of section 4.2 of ref.~\cite{UsingAlgebraicGeometry}),
a set of polynomials with these additional properties can be obtained as
\begin{equation}
e_i(z) = 1 - (1 - L_i(z)^\delta)^\delta \,,
\label{eq:partition-of-unity_polynomials}
\end{equation}
where $\delta$ is a positive integer such that for the intersection of the ideals
generated by each pole $J\langle\{ p_i \} \rangle = \langle z_1 - p_{i,1}, \ldots, z_n - p_{i,n} \rangle$
\begin{equation}
M \equiv \bigcap_{i=1}^m \langle z_1 - p_{i,1}, \ldots, z_n - p_{i,n} \rangle \,,
\label{eq:ideal_of_variety}
\end{equation}
we have that
\begin{equation}
M^\delta \subseteq I \,.
\label{eq:delta_definition}
\end{equation}
To find an appropriate $\delta$ thus requires algorithms to determine
the intersection and the product of two ideals and moreover to check
if one ideal is contained in another ideal.

To this end, we consider any two ideals in $\mathbb{C}[z_1, \ldots, z_n]$,
\begin{equation}
J_1 = \langle h_1, \ldots, h_r \rangle \hspace{5mm} \mathrm{and}
\hspace{5mm} J_2 = \langle k_1, \ldots, k_s \rangle \,.
\end{equation}
To compute the intersection $J_1 \cap J_2$, we introduce a parameter $t$
and consider the ideal
\begin{equation}
\langle t h_1, \ldots, t h_r, \hspace{1mm} (1-t) k_1, \ldots, (1-t) k_s \rangle \,.
\end{equation}
Now compute a Gr{\"o}bner basis $G$  of the latter ideal wrt.~lexicographic order
in which $t$ is greater than the $z_i$. The elements of $G$ which do not contain
the parameter $t$ will then form a basis\footnote{In fact, the basis will be
a Gr{\"o}bner basis of $J_1 \cap J_2$.}
of $J_1 \cap J_2$ (cf.~section 4.3 of ref.~\cite{IdealsVarietiesAlgorithms}).

The product of $J_1$ and $J_2$ is generated by the product of the generators,
\begin{equation}
J_1 J_2 \hspace{1mm}=\hspace{1mm} \langle h_i k_j \hspace{0.5mm} : \hspace{0.5mm} 1 \leq i \leq r \,, \hspace{2mm} 1 \leq j \leq s \rangle \,,
\end{equation}
cf.~proposition 6 of section 4.3 of ref.~\cite{IdealsVarietiesAlgorithms}.

Finally, to check the inclusion of ideals, for example whether $J_1 \subseteq J_2$,
compute a Gr{\"o}bner basis $H$ of $J_2$. Then
\begin{equation}
J_1 \subseteq J_2 \hspace{4mm} \Longleftrightarrow \hspace{4mm}
\forall \hspace{0.4mm} i = 1,\ldots, r \hspace{0.7mm} : \hspace{0.7mm} h_i \equiv 0 \hspace{3mm} \mbox{(mod $J_2$)} \,.
\end{equation}
That is, the inclusion $J_1 \subseteq J_2$ holds if and only if
all the generators of $J_1$ have a vanishing remainder upon polynomial division wrt.~$H$
(cf.~exercise 2 of section 1.4 of ref.~\cite{IdealsVarietiesAlgorithms}).

The computation of $\delta$ in eq.~\eqref{eq:partition-of-unity_polynomials}
following the above steps can in some cases be computationally intensive. This is especially
true in cases where $\delta$ is large and $M$ has many generators,
so that a large number of polynomial divisions must be carried out
in order to compute the generators of $M^j$ (where $j = 2, \ldots, \delta$)
in the intermediate stages.

Alternatively, to compute partition-of-unity polynomials,
we may compute the maximum pole multiplicity $d_\mathrm{max}$,
and use $\delta = d_\mathrm{max}$ in eq.~\eqref{eq:partition-of-unity_polynomials}.
Thus, we turn to explaining how to compute the multiplicities of the
poles $p_i$. To this end we consider a linear form $\ell$ of the type discussed above
eq.~\eqref{eq:Lagrangian_polynomials_generic_formula} with the
property of mapping all poles $p_i \in V$ to distinct values. We aim to find
the matrix of the map $P(z) \mapsto \ell(z) P(z)$ acting on polynomials in $Q$
and compute the dimensions of the eigenspaces of the matrix, as these are
the desired pole multiplicities (cf.~the discussion below Proposition (2.7) of
Chapter 4 of ref.~\cite{UsingAlgebraicGeometry}).

Let $\mathcal{C} = \big( c_1 (z), \ldots, c_D (z) \big)$ denote
the canonical basis of $Q$. To find the matrix $M_\ell$ of $P(z) \mapsto \ell(z) P(z)$,
for a fixed $1 \leq i \leq D$ decompose (the polynomial remainder of)
$\ell(z) c_i (z)$ in the basis $\mathcal{C}$,
producing a vector $v_i$ with $D$ entries. Then, cf.~section 2.4
of ref.~\cite{UsingAlgebraicGeometry}, $(M_\ell)_{ij} = (v_i)_j$.
The eigenvalues of $M_\ell$ are $(\lambda_1, \ldots, \lambda_m)
= \big( \ell(p_1), \ldots, \ell(p_m) \big)$, and the multiplicity of the
pole $p_i$ is the (algebraic) multiplicity $d_i$ of the eigenvalue $\ell(p_i)$,
\begin{equation}
\det (\lambda I - M_\ell) \hspace{0.8mm}=\hspace{0.8mm}
\big(\lambda - \ell(p_1)\big)^{d_1} \cdots \big(\lambda - \ell(p_m)\big)^{d_m} \,.
\end{equation}
In practice, the largest pole multiplicity $d_\mathrm{max} =
\mathrm{max} \{d_1, \ldots, d_m\}$ is strictly greater than
the smallest $\delta$ satisfying eq.~\eqref{eq:delta_definition}. As a result,
in cases where the dimension of the quotient ring is large, the
subsequent computation of the partition-of-unity polynomials
in eq.~\eqref{eq:partition-of-unity_polynomials} may prove time-consuming, even though the
computation of $\delta = d_\mathrm{max}$ itself is typically faster.

\subsection{Evaluation of residues}

With the construction of the canonical basis
$\mathcal{C} = \{ c_1(z), \ldots, c_D (z) \}$ and of the
corresponding dual basis $\mathcal{D} = \{ d_1(z), \ldots, d_D (z) \}$
of the quotient ring $Q$, along with the partition-of-unity polynomials
$e_i (z)$, all ingredients are now in place to compute the residues
of any given rational $n$-form,
\begin{equation}
\omega = \frac{h(z) \hspace{0.4mm} \d z_1 \wedge \cdots \wedge \d z_n}{P_1 (z) \cdots P_n (z)} \,.
\label{eq:diff_form}
\end{equation}
The key point is that the residue $\mathop{\mathrm{Res}}_I : Q \to \mathbb{C}$
defines a symmetric non-degenerate inner product on $Q$,
\begin{equation}
\langle h_1 \,, h_2 \rangle \hspace{0.5mm}\equiv\hspace{0.5mm} \mathop{\mathrm{Res}}_I (h_1 h_2) \,,
\end{equation}
and that $\mathcal{D}$ is dual to $\mathcal{C}$ with respect to this inner product,
\begin{equation}
\mathop{\mathrm{Res}}_I (c_a d_b) \hspace{1mm}=\hspace{1mm} \delta_{a b} \,,
\end{equation}
cf.~ref.~\cite{CattaniDickenstein}.

Thus, if we decompose the numerator $h(z)$ of eq.~\eqref{eq:diff_form} in the canonical basis,
\begin{equation}
h(z) = \lambda_1 c_1 (z) + \cdots + \lambda_D c_D (z) \,,
\end{equation}
and decompose the constant 1 in the dual basis,
\begin{equation}
1 = \mu_1 d_1 (z) + \cdots + \mu_D d_D (z) \,,
\end{equation}
then we can compute the global residue (cf.~eq.~\eqref{eq:global_residue_def})
of $\omega$ wrt.~$I$ as the dot product of the coefficient vectors,
\begin{align}
\mathop{\mathrm{Res}}_I \omega \hspace{0.7mm} &\equiv \hspace{0.7mm} \mathop{\mathrm{Res}}_I h(z)
= \mathop{\mathrm{Res}}_I \big( h(z) \cdot 1 \big)
= \mathop{\mathrm{Res}}_I \Big(\sum_{a=1}^D \lambda_a c_a \sum_{b=1}^D \mu_b d_b \Big) \\
&= \hspace{0.7mm} \sum_{a,b=1}^D \lambda_a \mu_b \hspace{0.6mm} \mathop{\mathrm{Res}}_I (c_a d_b)
\hspace{0.7mm}=\hspace{0.7mm} \sum_{a=1}^D \lambda_a \mu_a \,.
\end{align}
This prescription allows us to compute the global residue of $\omega$ wrt.~$I$,
i.e.~the sum of the residues at all poles in the associated variety
$V = \{ p_1, \ldots, p_m\}$. To compute the residue at any given pole $p_i$,
we utilize the corresponding partition-of-unity polynomial $e_i$,
\begin{equation}
\mathop{\mathrm{Res}}_{I, \hspace{0.6mm} z=p_i} \omega \hspace{0.6mm}=\hspace{0.6mm}
\mathop{\mathrm{Res}}_I (\omega e_i) \,.
\end{equation}

\section{Example of residue computation}\label{sec:example}

In this section we aim to apply the theory explained in
section~\ref{sec:quotient_ring_duality} to an example.
Thus, let us consider the differential form
\begin{equation}
\omega = \frac{(z_1 - z_2) \hspace{0.6mm} \d z_1 \wedge \d z_2}{z_1^2 (\chi  z_1 + 1)^2 z_2^3 (z_2-1)} \,,
\label{eq:diff_form_example}
\end{equation}
where $\chi$ is considered as a parameter. As $\omega$
depends on $n=2$ variables, we must partition
the denominator into two distinct factors, cf.~the discussion in
section~\ref{sec:General_theory}. We will consider the ideal
\begin{equation}
I \hspace{0.2mm}=\hspace{0.2mm} \langle P_1(z), P_2(z)\rangle
\hspace{0.4mm}=\hspace{0.4mm} \big\langle z_1^2 (z_2-1), \hspace{1mm} (\chi  z_1 + 1)^2 z_2^3 \big\rangle \,.
\label{eq:ideal_example}
\end{equation}
This has the associated variety
\begin{equation}
V = \{ p_1, p_2 \} = \left\{ (0,0), \big({-}\textstyle{\frac{1}{\chi}},1 \big) \right\} \,.
\label{eq:variety_example}
\end{equation}
As $V$ is finite, $I$ is zero-dimensional, so that $\omega$ has
well-defined residues at the poles in $V$.

\subsection{Computation of canonical basis of $Q$}
\begin{figure}[!h]
\begin{center}
\includegraphics[width=0.55\textwidth]{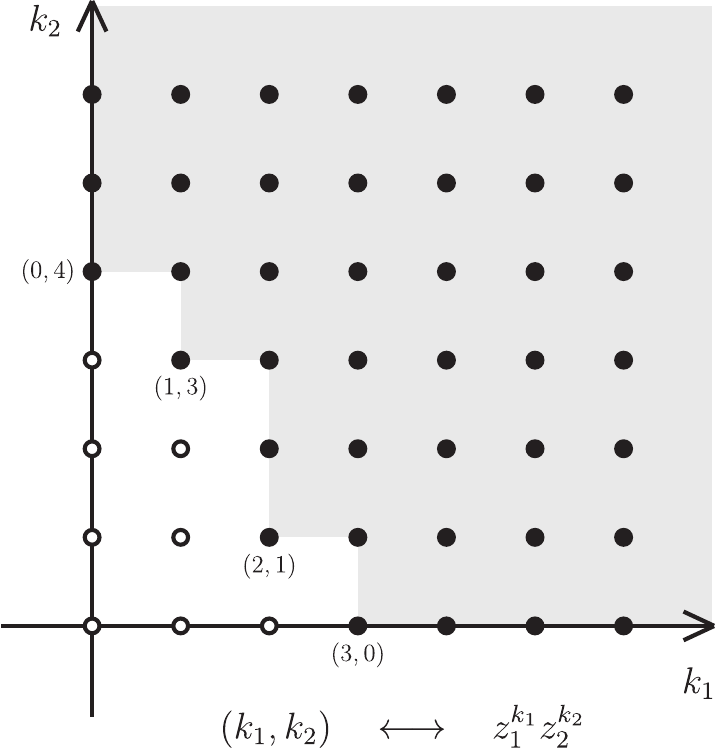}
\caption{The labeled lattice points are the exponent vectors
of the leading term of each element of the Gr{\"o}bner basis $G$
in eq.~\eqref{eq:Groebner_basis_example}. The lattice points
indicated with white circles are the elements of the set
$E$ defined in eq.~\eqref{eq:exponent_vectors_of_can_basis}.
These elements define exponent vectors of the canonical basis elements.}
\label{fig:canonical_basis}
\end{center}
\end{figure}

We choose as the monomial order $\prec$ degree lexicographic order.
The Gr{\"o}bner basis is found to be,\footnote{As $I$ has parameters,
the Gr{\"o}bner basis must be computed with the
\texttt{CoefficientDomain->RationalFunctions} option.}
\begin{equation}
G = \big\{ z_1^2 z_2 - z_1^2, \hspace{1.2mm}
-2 \chi^3 z_1^3 -3\chi^2 z_1^2 + z_2^3, \hspace{1.2mm}
z_2^4 - z_2^3, \hspace{1.2mm}
2\chi z_1 z_2^3 + \chi^2 z_1^2 + z_2^3 \big\} \,.
\label{eq:Groebner_basis_example}
\end{equation}
The leading terms of these elements are,
\begin{equation}
\mathrm{LT}(G) = \big\{ z_1^2 z_2, \hspace{1.2mm} -2 \chi ^3 z_1^3, \hspace{1.2mm} z_2^4, \hspace{1.2mm} 2\chi z_1 z_2^3 \big\} \,,
\end{equation}
whose corresponding exponent vectors are
\begin{equation}
\big\{ (2, 1), (3, 0), (0, 4), (1, 3) \big\} \,.
\end{equation}
From these exponent vectors we can now proceed to construct
the set $E$ defined in eq.~\eqref{eq:exponent_vectors_of_can_basis}.
In the case at hand, the construction is made more transparent
with the lattice illustration in fig.~\ref{fig:canonical_basis}.

Hence we find the following exponent vectors of the canonical
basis elements,
\begin{equation}
E = \big\{ (1,2), (0,3), (2,0), (1,1), (0,2), (1,0), (0,1), (0,0) \big\} \,.
\end{equation}
Thus, the canonical basis takes the following form,
\begin{equation}
\mathcal{C} = \big( z_1 z_2^2, z_2^3, z_1^2, z_1 z_2, z_2^2, z_1, z_2, 1 \big) \,.
\label{eq:canonical_basis_example}
\end{equation}
We remark that different choices of monomial order lead to
different canonical bases (but of course with the same number of elements).
For example, if we choose lexicographic monomial
order, we find the following canonical basis,
\begin{equation}
\mathcal{C}_\mathrm{lex} = \big( z_1 z_2^3, z_1 z_2^2, z_1 z_2, z_1, z_2^3, z_2^2, z_2, 1 \big) \,.
\end{equation}

\subsection{Computation of dual basis of $Q$}

To construct the dual basis of $Q$ wrt.~$\mathcal{C}$, we start by
computing the Bezoutian matrix of the ideal generators $P_i(z)$ in eq.~\eqref{eq:ideal_example}.
We find, cf.~eq.~\eqref{eq:Bezoutian_matrix_definition},
\begin{equation}
\mathrm{Bez} (z,y)
=\left( \hspace{-1mm}
\begin{array}{cc}
 (y_1+z_1) (z_2-1)                & y_1^2 \\[3mm]
 \chi  (\chi  (y_1+z_1)+2) z_2^3  & (\chi  y_1+1)^2 \left(y_2^2+z_2 y_2+z_2^2\right)
\end{array} \hspace{-1mm}
\right) \,.
\end{equation}
The Bezoutian determinant is thus,
\begin{align}
\mathcal{B}(z,y) &=
\chi^2 y_1^3 \left(y_2^2 (z_2-1)+y_2 (z_2-1) z_2-z_2^2\right) \nonumber \\
&\hspace{4mm} +y_1 (z_2{-}1) \left(y_2 z_2+y_2^2+z_2^2\right) (2 \chi z_1{+}1)+z_1 (z_2{-}1) \left(y_2 z_2+y_2^2+z_2^2\right)  \nonumber \\
&\hspace{4mm} +\chi y_1^2 \left(y_2^2 (z_2-1)+y_2 (z_2-1) z_2-z_2^2\right) (\chi z_1+2) \,.
\end{align}
We now perform polynomial division of $\mathcal{B}(z,y)$ with respect
to the Gr{\"o}bner basis $G$, where the elements are taken as
polynomials in the $z_i$ variables. This produces the following
decomposition, cf.~eq.~\eqref{eq:pol_div_of_Bez_wrt_z},
\begin{align}
(q_1, q_2, q_3, q_4) &= \big( 0,0,0, \textstyle{\frac{1}{2\chi}} + y_1 \big) \\
\mathcal{B}_Q(z,y) &= (2 \chi +\chi^2 y_1) \Big[y_1^2 \left(y_2^2 (z_2-1)+y_2 (z_2-1) z_2-z_2^2\right) \nonumber \\[-2mm]
& \hspace{28mm} +y_1 z_1 \left(y_2^2 (z_2-1)+y_2 (z_2-1) z_2-z_2^2\right)-z_1^2\Big] \nonumber \\
& \hspace{4mm} +(y_1+z_1) \left(y_2^2 (z_2-1)+y_2 (z_2-1) z_2-z_2^2\right)-\textstyle{\frac{z_2^3}{2 \chi}}
+\textstyle{\frac{3}{2}} \chi z_1^2 \,.
\end{align}
Next we perform polynomial division of the remainder $\mathcal{B}_Q(z,y)$ with respect to $G$,
where the elements are now taken as polynomials in the $y_i$ variables,
$P_i(z) \to P_i(y)$. This produces the decomposition in eq.~\eqref{eq:pol_div_of_Bez_wrt_y} with
\begin{equation}
(\widehat{q}_1, \widehat{q}_2, \widehat{q}_3, \widehat{q}_4)
= \Big( \chi (z_2-1) (y_2+z_2+1) \big(\chi  (y_1+z_1)+2 \big),
\hspace{1.2mm} \textstyle{\frac{1}{2\chi}}, \hspace{0.6mm} 0, \hspace{0.6mm} 0 \Big) \,,
\end{equation}
where the remainder is $\mathcal{B}_{Q \otimes Q} (z,y) = c_1 (z) d_1 (y) + \cdots + c_D (z) d_D (y)$,
with the dual basis elements taking the following form,
\begin{align}
\mathcal{D} &= \Big( (z_2{-}1) (2 \chi z_1{+}1), \hspace{1.5mm}
-\textstyle{\frac{1}{2 \chi}}, \hspace{1.5mm}
- \chi^2 z_1{-}\textstyle{\frac{\chi}{2}}, \hspace{1.5mm}
z_2 (z_2{-}1) (2 \chi z_1 {+} 1), \nonumber \\
& \hspace{8mm} z_1 (z_2{-}1), \hspace{1.5mm}
-\chi^2 z_1^2{-}2\chi z_1 z_2^2{-}z_2^2, \hspace{1.5mm}
z_1 z_2 (z_2{-}1), \hspace{1.5mm}
- z_1 z_2^2 {-} \textstyle{\frac{1}{2 \chi}} z_2^3 {-}\textstyle{\frac{\chi}{2}} z_1^2 \Big) \,,
\label{eq:dual_basis_example}
\end{align}
where we have expressed the elements as functions of the $z_i$ variables.

\subsection{Constructing partition-of-unity polynomials}

To construct the partition-of-unity polynomials
in eq.~\eqref{eq:partition-of-unity_polynomials}, our first aim
is to find a linear form which maps the elements $p_i$ of the
variety \eqref{eq:variety_example} to distinct values.
We observe that the linear form
\begin{equation}
\ell (z_1, z_2) = z_1 + z_2 \,,
\label{eq:linear_form_example}
\end{equation}
has this property. From eq.~\eqref{eq:Lagrangian_polynomials_generic_formula}
we then obtain the following Lagrange polynomials,
\begin{equation}
(L_1 (z), L_2 (z)) \hspace{1mm}=\hspace{1mm} \left( \frac{\chi (z_1+z_2-1)+1}{1-\chi}, \hspace{1.5mm} \frac{\chi (z_1+z_2)}{\chi-1} \right) \,,
\label{eq:Lagrange_polynomials_example}
\end{equation}
and we observe that $L_i (p_j) = \delta_{ij}$, as desired.

Our next aim is to compute $\delta$ as defined in eq.~\eqref{eq:delta_definition}.
To this end, we first compute $M$, defined in eq.~\eqref{eq:ideal_of_variety}
as the intersection of the ideals associated with each pole of the variety.
In the case at hand, the ideals associated with each pole of the variety \eqref{eq:variety_example}
are
\begin{align}
J\langle \{p_1\}\rangle &= \langle z_1, z_2 \rangle \\
J\langle \{p_2\}\rangle &= \left\langle z_1 + \textstyle{\frac{1}{\chi}}, z_2 -1 \right\rangle \,.
\end{align}
Accordingly, we introduce the parameter $t$ and compute the Gr{\"o}bner basis of
\begin{equation}
\big\langle t z_1, t z_2, (1-t) \big(z_1 + \textstyle{\frac{1}{\chi}} \big), (1-t)(z_2 -1) \big\rangle
\end{equation}
wrt.~lexicographic monomial order and the variable order $(t, z_1, z_2)$. We find
$\{ z_1+\frac{z_2}{\chi}, z_2^2-z_2, z_2 + t - 1 \}$. The elements which do not contain
the parameter $t$ then form a basis of $J\langle \{p_1\}\rangle \cap J\langle \{p_2\}\rangle$,
\begin{equation}
M = \big\langle z_1 + \textstyle{\frac{z_2}{\chi}}, z_2^2-z_2 \big\rangle \,.
\label{eq:ideal_of_M_example}
\end{equation}
Now, to determine $\delta$, we start by checking whether $M \subseteq I$.
Polynomial division of the elements of eq.~\eqref{eq:ideal_of_M_example}
wrt.~the Gr{\"o}bner basis of $I$ in eq.~\eqref{eq:Groebner_basis_example}
leaves remainders identical to the original elements.

Thus, we proceed to consider $M^2$. Taking the products of the generators
in eq.~\eqref{eq:ideal_of_M_example} and performing polynomial division
we find
\begin{equation}
M^2 = \big\langle \big( z_1 +\textstyle{\frac{z_2}{\chi}} \big)^2, \hspace{1mm}
(z_2^2-z_2) \big(z_1 +\textstyle{\frac{z_2}{\chi}} \big), \hspace{1mm}
z_2^2-z_2^3 \big\rangle \,.
\label{eq:ideal_of_M2_example}
\end{equation}
As the remainders are non-zero, we proceed to consider $M^3$. Taking the
products of the generators in eqs.~\eqref{eq:ideal_of_M_example} and \eqref{eq:ideal_of_M2_example}
and performing polynomial division we find
\begin{align}
M^3 &= \Big\langle \textstyle{\frac{3}{2}} \left( \textstyle{\frac{z_2^3}{\chi^3}}+\textstyle{\frac{2 z_1 z_2^2}{\chi^2}}+\textstyle{\frac{z_1^2}{\chi}}\right), \hspace{1.5mm}
-\textstyle{\frac{z_2^3}{\chi^2}}-\textstyle{\frac{2 z_1 z_2^2}{\chi}}-z_1^2, \hspace{1.5mm}
-\textstyle{\frac{z_2^3}{\chi^2}}-\textstyle{\frac{2 z_1 z_2^2}{\chi}}-z_1^2, \hspace{1.5mm} \nonumber \\
& \hspace{8.5mm} \textstyle{\frac{1}{2}} \left(\textstyle{\frac{z_2^3}{\chi}}+\chi z_1^2+2 z_1 z_2^2\right), \hspace{1.5mm}
\textstyle{\frac{1}{2}} \left(\textstyle{\frac{z_2^3}{\chi}}+\chi z_1^2+2 z_1 z_2^2\right), \hspace{1.5mm}
0 \Big\rangle \,.
\label{eq:ideal_of_M3_example}
\end{align}
As the remainders are non-zero, we proceed to consider $M^4$. Taking the
products of the generators in eqs.~\eqref{eq:ideal_of_M_example} and \eqref{eq:ideal_of_M3_example}
and performing polynomial division we find
\begin{equation}
M^4 = \langle 0 \rangle \,.
\label{eq:ideal_of_M4_example}
\end{equation}
Hence we conclude that $\delta = 4$.

Alternatively, we may use the maximum pole multiplicity as a
value for $\delta$. To this end we compute the matrix
$M_\ell$ of $P(z) \mapsto \ell(z) P(z)$ for the linear form
in eq.~\eqref{eq:linear_form_example}. Using the canonical basis
$\mathcal{C} = (c_1(z), \ldots, c_8(z))$ in eq.~\eqref{eq:canonical_basis_example},
we find
\begin{equation}
M_\ell
\hspace{0.8mm}=\hspace{0.8mm} \left(
\begin{array}{cccccccc}
 0 & -\frac{1}{2\chi} & 1{-}\frac{\chi}{2} & 0 & 0 & 0 & 0 & 0 \\[1mm]
 0 & 1{-}\frac{1}{2\chi} & -\frac{\chi}{2} & 0 & 0 & 0 & 0 & 0 \\[1mm]
 0 & \frac{1}{2\chi^3} & 1{-}\frac{3}{2\chi} & 0 & 0 & 0 & 0 & 0 \\[1mm]
 1 & 0 & 1 & 0 & 0 & 0 & 0 & 0 \\
 1 & 1 & 0 & 0 & 0 & 0 & 0 & 0 \\
 0 & 0 & 1 & 1 & 0 & 0 & 0 & 0 \\
 0 & 0 & 0 & 1 & 1 & 0 & 0 & 0 \\
 0 & 0 & 0 & 0 & 0 & 1 & 1 & 0 \\
\end{array}
\right) \,,
\end{equation}
i.e., so that $\ell(z) c_i (z) \equiv \sum_{j=1}^8 (M_\ell)_{ij} c_j (z) \mbox{ (mod $I$)}$.
From this we find
\begin{equation}
\det (\lambda I - M_\ell) \hspace{0.8mm}=\hspace{0.8mm} \big(\lambda - \ell(p_1)\big)^6 \big(\lambda - \ell(p_2)\big)^2 \,.
\label{eq:eigenvalues_of_M_l_example}
\end{equation}
We conclude that the poles $p_1$ and $p_2$ have the multiplicities 6 and 2,
respectively. In particular, $d_\mathrm{max} = 6$.

Plugging the value $\delta=4$ from eq.~\eqref{eq:ideal_of_M4_example} into
eq.~\eqref{eq:partition-of-unity_polynomials}
with the Lagrange polynomials given in eq.~\eqref{eq:Lagrange_polynomials_example}
and performing polynomial division wrt.~$G$,%
\footnote{To calculate a desired power of a polynomial $L_i(z)^\delta$ in the quotient ring $Q$,
\textsc{MultivariateResidues} performs polynomial division wrt.~$G$ after
taking each product $L_i^k = L_i^{k-1} L_i$. This guarantees that each power computed
in the intermediate stages has $D = \mathrm{dim} \hspace{0.7mm} Q$ terms
rather than $D^k$ terms, thereby minimizing intermediate expression swell.} we find%
\footnote{The value $\delta=d_\mathrm{max}=6$ obtained from eq.~\eqref{eq:eigenvalues_of_M_l_example}
produces identical results for $e_1(z)$ and $e_2(z)$.}
\begin{equation}
e_1(z) = 1 - z_2^3 \hspace{6mm} \mathrm{and} \hspace{6mm} e_2(z) = z_2^3 \,.
\label{eq:partition-of-unity_polynomials_example}
\end{equation}
It is straightforward to check that these polynomials indeed have the
properties stated in eqs.~\eqref{eq:partition-of-unity_property_1}--\eqref{eq:partition-of-unity_property_2}.

Now, to compute the residues at each pole $p_i$ in the variety \eqref{eq:variety_example},
we utilize the partition-of-unity polynomials computed in
eq.~\eqref{eq:partition-of-unity_polynomials_example} and
consider the numerator of eq.~\eqref{eq:diff_form_example}, i.e.~$h(z) = z_1 - z_2$,
multiplied by these polynomials,
\begin{equation}
\begin{alignedat}{4}
  & h(z) e_1 (z)  && = (z_1 - z_2) (1 - z_2^3)  && \hspace{1mm} \equiv \textstyle{\frac{(2\chi+1)}{2\chi}} z_2^3
+ \textstyle{\frac{\chi}{2}} z_1^2 + z_1 - z_2  && \hspace{6mm} \mbox{(mod $I$)} \,, \\
  & h(z) e_2 (z)  && = (z_1 - z_2) z_2^3        && \hspace{1mm} \equiv -\textstyle{\frac{2\chi+1}{2\chi}} z_2^3
-\textstyle{\frac{\chi}{2}} z_1^2               && \hspace{6mm} \mbox{(mod $I$)} \,.
\end{alignedat}
\end{equation}
We proceed to decompose these in the canonical basis $\mathcal{C}$
given in eq.~\eqref{eq:canonical_basis_example}, finding the following coefficient
vectors,
\begin{align}
\Lambda_1 &= \big( 0, \textstyle{\frac{2\chi+1}{2\chi}},   \frac{\chi}{2},0,0,1,-1,0 \big) \,,\\
\Lambda_2 &= \big( 0, -\textstyle{\frac{2\chi+1}{2\chi}}, -\frac{\chi}{2},0,0,0,0,0 \big) \,,
\end{align}
i.e., where $h(z) e_i (z) \equiv \Lambda_i \cdot \mathcal{C}$ (mod $I$). Moreover,
the constant 1 may be composed in the dual basis $\mathcal{D}$
in eq.~\eqref{eq:dual_basis_example} as $1 = \mu \cdot \mathcal{D}$ where,
\begin{equation}
\mu = (0, -2\chi,0,0,0,0,0,0) \,.
\end{equation}
Thus we find for the residues,
\begin{equation}
\begin{alignedat}{4}
  & \mathop{\mathrm{Res}}_{I, \hspace{0.6mm} z=p_1} \omega && \hspace{1mm}=\hspace{1mm} \Lambda_1 \cdot \mu  && \hspace{1mm}=\hspace{1mm} -2 \chi -1 \,, \\
  & \mathop{\mathrm{Res}}_{I, \hspace{0.6mm} z=p_2} \omega && \hspace{1mm}=\hspace{1mm} \Lambda_2 \cdot \mu  && \hspace{1mm}=\hspace{1mm} 2 \chi +1 \,.
\end{alignedat}
\end{equation}

\section{Global residue theorems}\label{sec:global_residue_thm}

In this section we discuss global residue theorems for multivariate meromorphic
forms. In the univariate case it is well known that the sum of all residues,
including that at infinity, equals zero,
\begin{equation}
\sum_{i=1}^m \mathop{\mathrm{Res}}_{z=p_i} \varpi \hspace{1mm}=\hspace{1mm} 0 \,,
\end{equation}
where $\{p_1, \ldots, p_m\} \subset \mathbb{CP}^1$ denote the poles of $\varpi$.

This property generalizes to the multivariate case,
\begin{align}
\omega = \frac{h(z) \hspace{0.4mm} \d z_1\wedge\cdots\wedge \d z_n}{f_1(z)\cdots f_n(z)}\,,
\label{eq:diff_form_on_open_cover}
\end{align}
where, however, one typically has several linear relations which relate the residues
at finite locations to residues at infinity. The existence of these relations
follows from the following theorem.

\clearpage

\begin{theorem}\label{thm:global_residue_thm}
(Global residue theorem). Let $\omega$ denote a meromorphic $n$-form defined
on a compact manifold $M$. Given an open covering $\{ U_i \}$, let $\omega$ take the local
form given in eq.~\textup{(\ref{eq:diff_form_on_open_cover})}.
Furthermore, let $D_j = \{ z \in M : f_j (z) = 0\}$ with $j=1,\ldots, n$
denote the divisors of $\omega$, and assume that $V = D_1 \cap \cdots \cap D_n$
is a finite set. Then
\begin{equation}
\sum_{p \in V} \mathop{\mathrm{Res}}_{p} \omega \hspace{1mm}=\hspace{1mm} 0 \,,
\label{eq:global_residue_thm}
\end{equation}
where each $\mathop{\mathrm{Res}}_{p} \omega$ is evaluated locally on a patch $U_i$
which contains $p$.
\end{theorem}

For a proof we refer to section 5.1 of ref.~\cite{GriffithsHarris}.

As the global residue theorem applies to forms defined on compact manifolds,
in order to apply it to an $n$-form defined on $\mathbb{C}^n$,
one must add a boundary at infinity. One convenient way to do so is to embed
$\mathbb{C}^n$ in complex projective space $\mathbb{CP}^n$, which is compact.
In the following we will show how to apply the global residue theorem
for this choice of compactification. However, we emphasize that other
compactifications exist, corresponding to alternative ways of adding
a boundary at infinity, and will lead to different residue relations
produced by the global residue theorem.

We recall that $\mathbb{CP}^n$ can be defined as the space of $(n+1)$-tuples of
complex numbers $W = (w_0, \ldots, w_n) \in \mathbb{C}^{n+1}\setminus\{0\}$
where two elements are identified if they lie along the same line passing through
the origin, $tW \sim W$ for $t\in \mathbb{C}\setminus\{ 0 \}$. That is,
\begin{equation}
\mathbb{CP}^n = \{ (w_0,\ldots,w_n) \neq 0 \} \Big/ \{ tW \sim W  \hspace{2mm} \mathrm{where} \hspace{2mm} t \neq 0\} \,.
\end{equation}
A covering of $\mathbb{CP}^n$ is provided by the patches
\begin{equation}
U_i = \{ (w_0,\ldots,w_n) \hspace{0.8mm} : \hspace{0.8mm} w_i = 1 \}
\hspace{4mm} \mathrm{where} \hspace{4mm} i = 0, \ldots, n \,.
\label{eq:patches_on_CPn}
\end{equation}
Here $\mathbb{C}^n = \{ (z_1,\ldots,z_n) \}$ can be identified with the patch
$U_0$ by using the homogeneous coordinates,
\begin{equation}
z_1 = \frac{w_1}{w_0}\,, \hspace{3mm} \ldots, \hspace{3mm} z_n = \frac{w_n}{w_0}\,,
\label{eq:homogeneous_coords}
\end{equation}
since on patch $U_0$ we have $w_0 = 1$. Thus, $\mathbb{C}^n \subset \mathbb{CP}^n$.
Points with $w_0 = 0$ are referred to as \emph{points at infinity}.
For the Riemann sphere $\mathbb{CP}^1$, the patches $U_0$ and $U_1$ are
the Riemann sphere with respectively the point at infinity $(0,1)$,
and the origin $(1,0)$, removed.

To define the differential form in eq.~(\ref{eq:diff_form_on_open_cover})
on each of the patches $U_k$ in eq.~(\ref{eq:patches_on_CPn}) we must find the
Jacobian from the coordinates $(z_1, \ldots, z_n)$ to $(w_0, \ldots, w_{k-1}, w_{k+1}, \ldots, w_n)$.
Using the homogeneous coordinates in eq.~(\ref{eq:homogeneous_coords}), it is straightforward to show that
\begin{equation}
\mathop{\mathrm{det}}_{\substack{i\in \{1,\ldots,n\} \phantom{\setminus\{ k \}} \\ j \in \{ 0, \ldots, n\}\setminus\{ k \} }}
\frac{\partial z_i}{\partial w_j} \hspace{2mm}=\hspace{2mm} \frac{(-1)^k}{w_0^{n+1}} \,.
\end{equation}
Letting $\widehat{\d w_k}$ denote that the respective differential has been dropped,
we therefore find that $\omega$ evaluated on the patch $U_k$ takes the form,
\begin{align}
\omega \big|_{U_k} \hspace{1mm}=\hspace{1mm} \frac{(-1)^k \hspace{0.3mm} h\big(\textstyle{\frac{w}{w_0}}\big)
\hspace{0.4mm} \d w_0\wedge\cdots\wedge \widehat{\d w_k} \wedge\cdots\wedge \d w_n}
{w_0^{n+1} f_1\big(\textstyle{\frac{w}{w_0}}\big)\cdots f_n\big(\textstyle{\frac{w}{w_0}}\big)}\,.
\end{align}
To apply the global residue theorem, we must then consider
each partition of the factors contained in the set
$\{w_0^{n+1}, f_1\big(\textstyle{\frac{w}{w_0}}\big),$ $\ldots, f_n\big(\textstyle{\frac{w}{w_0}}\big)\}$
into $n$ divisors. Each partition gives rise to a linear relation,
as we will see in the following example.

\subsection{Example: application of the global residue theorem}\label{sec:GRT_example}

To illustrate how the global residue theorem (theorem~\ref{thm:global_residue_thm})
yields linear relations between the residues of a meromorphic form,
we consider as an example the form $\omega$ given in eq.~(\ref{eq:degenerate_diff_form}).
Expressed in terms of the homogeneous coordinates (\ref{eq:homogeneous_coords}),
$\omega$ takes the following form on patch $U_k$,
\begin{align}
\omega \big|_{U_k} \hspace{1mm}=\hspace{1mm}
\frac{(-1)^k \hspace{0.4mm} w_1 \hspace{0.6mm}
\d w_0 \wedge \cdots \wedge \widehat{\d w_k} \wedge \cdots \wedge \d w_2}
{w_0 w_2 (a_1 w_1 + a_2 w_2)(b_1 w_1 + b_2 w_2)} \,.
\end{align}
In this case there are seven distinct partitions of the
four denominator factors. Let us consider the following partition,
\begin{equation}
f_1(w) = w_0 w_2 \,, \hspace{4mm} f_2(w) = (a_1 w_1 + a_2 w_2)(b_1 w_1 + b_2 w_2) \,,
\end{equation}
giving rise to the divisors $D_i = \{(w_0, w_1, w_2) \hspace{0.3mm} : \hspace{0.3mm} f_i(w) = 0\}$.
We find that the intersection of the divisors is a finite set,
\begin{equation}
V = D_1 \cap D_2 = (p_1, p_2, p_3) = \Big( (1,0,0), \big(0,1,-\textstyle{\frac{a_1}{a_2}}\big),
\big(0,1,-\textstyle{\frac{b_1}{b_2}}\big)\Big) \,,
\end{equation}
and hence the global residue theorem applies. Noting that $p_1 \in U_0$ and $p_{2,3} \in U_1$,
we can evaluate the residues on these respective patches, dropping the constant $w_0$ and
$w_1$ entries, respectively. For the left-hand side of eq.~(\ref{eq:global_residue_thm}) we then find,
\begin{align}
\sum_{p \in V} \mathop{\mathrm{Res}}_{p} \omega &=
\mathop{\mathrm{Res}}_{I_0; \hspace{0.3mm} p_1} \frac{w_1 \hspace{0.6mm} \d w_1\wedge \d w_2}
{w_2 (a_1 w_1 {+} a_2 w_2)(b_1 w_1 {+} b_2 w_2)} +
\mathop{\mathrm{Res}}_{I_1; \hspace{0.3mm} p_2} \frac{(-1) \hspace{0.6mm}
\d w_0\wedge \d w_2}{w_0 w_2 (a_1 {+} a_2 w_2)(b_1 {+} b_2 w_2)} \nonumber \\
& \hspace{5mm} + \mathop{\mathrm{Res}}_{I_1; \hspace{0.3mm} p_3} \frac{(-1) \hspace{0.6mm}
\d w_0\wedge \d w_2}{w_0 w_2 (a_1 {+} a_2 w_2)(b_1 {+} b_2 w_2)} \,,
\end{align}
where the residues are computed with respect to the ideals
$I_j \equiv \big\langle f_1(w), f_2(w) \big\rangle\big|_{w_j = 1}$ for $j = 0,1$.
Explicitly, we find
\begin{align}
\sum_{p \in V} \mathop{\mathrm{Res}}_{p} \omega
\hspace{1mm}=\hspace{1mm} -\frac{1}{a_1 b_1} - \frac{a_2}{a_1 (a_1 b_2 - a_2 b_1)} - \frac{b_2}{b_1 (a_2 b_1 - a_1 b_2)}
\hspace{1mm}=\hspace{1mm} 0 \,,
\end{align}
in agreement with eq.~(\ref{eq:global_residue_thm}).

Analogously, for the partition
\begin{equation}
f_1(w) = w_0 (a_1 w_1 + a_2 w_2) \,, \hspace{4mm} f_2(w) = w_2 (b_1 w_1 + b_2 w_2) \,,
\end{equation}
we have
\begin{equation}
V = \Big( (1,0,0), \big(0,1,0\big), \big(0,1,-\textstyle{\frac{b_1}{b_2}}\big)\Big) \,,
\end{equation}
and obtain the residue relation,
\begin{align}
\sum_{p \in V} \mathop{\mathrm{Res}}_{p} \omega
\hspace{1mm}=\hspace{1mm} - \frac{a_2}{a_1 (a_1 b_2 - a_2 b_1)} -\frac{1}{a_1 b_1} - \frac{b_2}{b_1 (a_2 b_1 - a_1 b_2)}
\hspace{1mm}=\hspace{1mm} 0 \,,
\end{align}
again in agreement with eq.~(\ref{eq:global_residue_thm}).

The global residue theorems associated with the remaining five partitions
of the denominator factors of $\omega$ are computed analogously.

\section{Applications of multivariate residues}\label{sec:applications}

In this section we give two applications of multivariate residues, namely
the computation of generalized-unitarity cuts and the Cachazo-He-Yuan scattering
equations. The former are important in several contexts,
for example unitarity calculations of loop amplitudes and the Grassmannian
formulation of the S-matrix by Arkani-Hamed et al.

First we focus on unitarity calculations. To provide some context, let us
consider a one-loop scattering amplitude in a generic gauge theory. From standard
reduction techniques it can be shown that there is a finite basis of one-loop
integrals\footnote{To avoid any confusion, we emphasize that the above statement
is that the basis of one-loop integrals is a finite set. On the other hand, the
one-loop integrals themselves have ultraviolet and/or infrared divergences and
must be regulated, for example by use of dimensional regularization.} in which
the amplitude can be expanded \cite{Weinzierl:2006qs,Bern:2007dw,Ellis:2011cr},
\begin{equation}
A^{(1)}_n = \sum_{\mathrm{boxes}} c_\Box I_\Box
+ \sum_{\mathrm{triangles}} c_\triangle I_\triangle
+ \sum_{\mathrm{bubbles}} c_\circ I_\circ
+ \sum_{\mathrm{tadpoles}} c_\multimap I_\multimap +
\mbox{rational terms}\,, \label{eq:one-loop_basis_decomposition}
\end{equation}
where $I_\Box, I_\triangle, I_\circ$ and $I_\multimap$ represent
box, triangle, bubble and tadpole integrals, respectively. As all these integrals
are known \cite{Bern:1994cg}, this decomposition reduces the computation of $A^{(1)}_n$
to the computation of the expansion coefficients.

The coefficient $c_\Box$ of the box integral
\begin{equation}
I_\Box = \int_{\mathbb{R}^D} \frac{\d^D \ell}{(2\pi)^D}
\frac{1}{\prod_{i=1}^4 p_i^2 (\ell)} \,,
\end{equation}
can now be computed by replacing the integration contour $\mathbb{R}^D$ in
eq.~\eqref{eq:one-loop_basis_decomposition} by the contour
\begin{equation}
T^4_\epsilon = \{  \ell \in \mathbb{C}^4 : |p^2_i(\ell)|=\epsilon_i, \hspace{2.5mm} i=1,\ldots, 4 \} \, .
\label{eq:quad_cut_torus_formal}
\end{equation}
The replacement of contour $\mathbb{R}^D \to T^4_\epsilon$ has the
effect of computing the residue at the poles where all four propagators
$p_i^2(\ell)$ of $I_\Box$ go on shell. All terms missing
any one of the propagators therefore vanish, and eq.~\eqref{eq:one-loop_basis_decomposition}
becomes~\cite{Britto:2004nc}
\begin{equation}
c_\Box = \frac{1}{2}
\sum_{a\in \{ L,L^\bullet\}} \sum_{{\mathrm{helicities,}\atop \mathrm{species}}}
\prod_{i=1}^4 \hspace{0.4mm} A^\mathrm{tree}_i \big(p_i (a), p_{i+1} (a) \big) \,,
\label{eq:BCF_box_formula}
\end{equation}
where $L,L^\bullet$ denote the solutions to $p_i^2(\ell)=0$ for $i=1,\ldots,4$,
and $A^\mathrm{tree}_i$ are the tree amplitudes arising from literally cutting the
propagators of the box graph.

\begin{figure}[!h]
\begin{center}
\includegraphics[width=0.4\textwidth]{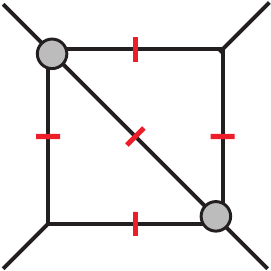}
\caption{(Color online.) A generalized-unitarity cut where the propagators with superimposed
red lines have been put on shell. The gray blobs represent tree amplitudes.}
\label{fig:cut_slashed_box}
\end{center}
\end{figure}

More to the point, though the integral $\int_{T^4_\epsilon} \frac{\d^D \ell}{(2\pi)^D}
\frac{1}{\prod_{i=1}^4 p_i^2 (\ell)}$ is a multivariate residue, it is non-degenerate
and can thus be computed directly from eq.~\eqref{eq:residue_from_Jacobian}.
However, starting at two loops, degenerate residues are generic, and the algorithms explained in
sections~\ref{sec:residues_from_transformation_formula}~and~\ref{sec:quotient_ring_duality}
become necessary to evaluate them. To give an explicit example, we consider the
generalized-unitarity cut shown in figure~\ref{fig:cut_slashed_box}.

It turns out that the five on-shell constraints can be integrated out as a
non-degenerate residue. However, the residues of the resulting integrand in
the remaining $2 \times 4 - 5 = 3$ variables will generically be degenerate: for example,
the residue with respect to the ideal $I$ (where $\chi$ denotes the ratio $t/s$
of Mandelstam invariants),
\begin{equation}
I = \langle z_1^2, \hspace{0.7mm} z_2, \hspace{0.7mm} -\chi z_1+\chi z_3 z_1+z_3 z_1+z_2 z_3-z_3 \rangle \,,
\end{equation}
at $(z_1,z_2,z_3)= (0,0,0)$ is degenerate, and the algorithms explained in
sections~\ref{sec:residues_from_transformation_formula}~and~\ref{sec:quotient_ring_duality}
are required to compute the residue.

Multivariate residues also play a central role in the Cachazo-He-Yuan scattering equations which describe tree-level
scattering amplitudes in any spacetime dimension \cite{Cachazo:2013hca}.
In this formalism, the scattering of $n$ particles with momenta $p_a$ is encoded
in the solutions $\sigma_a \in \mathbb{CP}^1$ to the scattering equations,
\begin{equation}
\sum_{b \neq a} \frac{s_{ab}}{\sigma_a - \sigma_b} = 0 \,, \label{eq:CHY_scattering_eqs}
\end{equation}
where $s_{ab} = (p_a + p_b)^2$ denote Mandelstam invariants. If the solutions
to these equations can be found, the tree-level amplitude in any dimension
can then be computed as,
\begin{equation}
A_n = \int \frac{\d^n \sigma}{\mathrm{vol} \hspace{0.6mm} SL(2,\mathbb{C})}
\prod_a{}' \hspace{0.9mm} \delta \hspace{-0.5mm} \left(\textstyle{\sum_{b\neq a} \frac{s_{ab}}{\sigma_{ab}}}\right)
\frac{E_n(\{p,\epsilon,\sigma\})}{\sigma_{12} \sigma_{23} \cdots \sigma_{n1}} \,,
\label{eq:CHY_form_of_amplitude}
\end{equation}
where $\sigma_{ij} \equiv \sigma_i - \sigma_j$, $\epsilon$ denotes polarization vectors,
and the prime on the product sign indicates that (any) three of the equations are
dropped (as only $n-3$ equations are independent). The quantity $E_n$
depends on the underlying theory (for Yang-Mills theory,
it is the Pfaffian of an antisymmetric matrix composed of all dot products of
$k_i$ and $\epsilon_j$, divided by $\sigma_{ab}$). The delta functions are
more properly understood as contour integrals.

In practice, obtaining the $(n-3)!$ solutions to eq.~(\ref{eq:CHY_scattering_eqs})
is difficult, and summing over the individual residues at each pole $\sigma_{ab}$ in
eq.~(\ref{eq:CHY_form_of_amplitude}) is not straightforward for high multiplicities $n$.
However, as observed in refs.~\cite{Sogaard:2015dba,Bosma:2016ttj}, these steps can
be circumvented by directly computing the global residue (\ref{eq:global_residue_def}).
In practice this leads to an efficient method for computing tree amplitudes
in the Cachazo-He-Yuan formalism.\footnote{It is worth pointing out that for $n \geq 5$,
the solutions to eq.~(\ref{eq:CHY_scattering_eqs}) and the individual residues
are \emph{algebraic} rather than rational in $s_{ab}$. Upon summing over all individual residues
eq.~(\ref{eq:CHY_form_of_amplitude}) the spurious radicals cancel, leaving a rational expression.
In contrast, the global residue is manifestly rational, as the integrand of eq.~(\ref{eq:CHY_form_of_amplitude})
is rational.}

\textsc{MultivariateResidues} allows direct computation of the global residue.
As a result, the package enables direct evaluation of tree amplitudes
in the Cachazo-He-Yuan formalism, bypassing the steps of solving eq.~(\ref{eq:CHY_scattering_eqs})
and summing over the individual residues. In section~\ref{sec:CHY_example}
we illustrate this by computing the five-scalar tree amplitude in $\phi^3$ theory.

We remark that multivariate residues also play an important role
in elimination theory in the context of solving systems of
multivariate polynomial equations \cite{CattaniDickenstein}.

\section{Manual}\label{sec:manual}

The package MultivariateResidues.m can be obtained from ref.~\cite{webpage}.
At the beginning of a Mathematica session, the package can be loaded with
\begin{CodeSample}
SetDirectory[NotebookDirectory[]];
<< "MultivariateResidues`"
\end{CodeSample}
where it is assumed that the package and the notebook are located in the same directory.
The newly available definitions can be shown by running
\begin{CodeSample}
?MultivariateResidues`*
\end{CodeSample}
The package defines one new function, called \Math{MultivariateResidue}.
Below we give a brief introduction to this function and its options.

\subsection{New functions}
\label{sec:newfunction}

The function \Math{MultivariateResidue} computes a multivariate residue, based on the algorithms described in this paper.
It has the following syntax:
\begin{CodeSample}
MultivariateResidue[Num, {d[1],..., d[n]},
					   {z[1] -> z[1,1], ..., z[n] -> z[n,1]}]
\end{CodeSample}
which returns the multivariate residue of $\mathrm{Num} / (d_1\,d_2 \dotsm d_n)$ at the location given by $(z_1,\dotsc,z_n) = (z_1^{(1)},\dotsc,z_n^{(1)})$.
Alternatively,
\begin{CodeSample}
MultivariateResidue[Num, {d[1],..., d[n]}, {z[1],...,z[n]},
				 {{z[1,1], ..., z[n,1]}, {z[1,2], ..., z[n,2]}, ...}]
\end{CodeSample}
returns a list of multivariate residues of $\mathrm{Num} / (d_1\,d_2 \dotsm d_n)$ at the collection of points $(z_1,\dotsc,z_n) \in \{ (z_1^{(1)},\dotsc,z_n^{(1)}), (z_1^{(2)},\dotsc,z_n^{(2)}), \dotsc \}$.
This second syntax is better suited for the computation of several residues, because it exploits the fact that part of the computation is common to all poles.

In the univariate case, \Math{MultivariateResidue} is equivalent to the native Mathematica function \Math{Residue}. For instance,
\begin{CodeSample}
MultivariateResidue[f[z], {z}, {z -> 0}]
Residue[f[z]/z, {z, 0}]
Out: f[0]
Out: f[0]
\end{CodeSample}

As a multivariate example, let us compute the residues considered in section \ref{sec:residues_from_transformation_formula}.
Taking $\varphi_1, \varphi_2, \varphi_3$ from eq.~\eqref{eq:denominator_factors_example}, we can straightforwardly compute the residues with respect to the ideals given in eq.~\eqref{eq:denominator_partitionings_example} as follows:
\begin{CodeSample}
f[1] = z[2];
f[2] = a[1] z[1] + a[2] z[2];
f[3] = b[1] z[1] + b[2] z[2];
MultivariateResidue[z[1], {f[1], f[2]*f[3]}, {z[1] -> 0, z[2] -> 0}]
MultivariateResidue[z[1], {f[2], f[3]*f[1]}, {z[1] -> 0, z[2] -> 0}]
MultivariateResidue[z[1], {f[3], f[1]*f[2]}, {z[1] -> 0, z[2] -> 0}]
Out: -(1/(a[1] b[1]))
Out: -(a[2]/(a[1] (-a[2] b[1] + a[1] b[2])))
Out: -(b[2]/(b[1] (a[2] b[1] - a[1] b[2])))
\end{CodeSample}

When using the second syntax, the input list of poles does not have to contain all points in the variety defined by the ideal (i.e., the set of points where all denominator factors vanish). Moreover, it may contain additional points (although the corresponding residues will be zero). As an example, consider the following ideal,
\begin{CodeSample}
Ideal = {z[1]^2 (1 + z[1] - z[2]), z[3]^3,
   		z[2]^3 (-1 - z[1] - a z[1] + z[2] + a z[1] z[3])};
Parameters = {a};
Vars = {z[1], z[2], z[3]};
\end{CodeSample}
The resulting variety contains three points,
\begin{CodeSample}
Variety = Map[Last, Sort[Solve[Ideal == 0, Vars]], {2}]
Out: {{-1, 0, 0}, {0, 0, 0}, {0, 1, 0}}
\end{CodeSample}
We may ask for the residue computed at precisely the poles contained in the variety, or alternatively a subset, or alternatively a subset with an added random point,
\begin{CodeSample}
MultivariateResidue[1, Ideal, Vars, Variety]
MultivariateResidue[1, Ideal, Vars, Variety[[1 ;; 2]]]
MultivariateResidue[1, Ideal, Vars, Join[Variety[[1 ;; 2]], {{0, 0, 1}}]]
Out: {-(6/a), 0, 6/a}
Out: {-(6/a), 0}
Out: {-(6/a), 0, 0}
\end{CodeSample}
In fact, the algorithm based on the dual structure of the quotient ring requires the computation of the residues of all poles in the variety. This particular algorithm will therefore internally compute the variety itself and return the residues at the (sub)set of poles that are specified by the user.

When using the method \Math{"QuotientRingDuality"} one may also specify \Math{Variety = \{GlobalResidue\}}. In this case the program returns the global residue (\ref{eq:global_residue_def}), which equals the sum of all local residues. The benefit is that the global residue can be calculated directly without determining the partition-of-unity polynomials $e_i$ in eq.~(\ref{eq:partition-of-unity_property_2}) required to compute the local residues. We refer to
section~\ref{sec:CHY_example} for an example.

\textsc{MultivariateResidues} contains an implementation of
the global residue theorem discussed in section~\ref{sec:global_residue_thm},
choosing to extend the input differential form $\omega$
to complex projective space $\mathbb{CP}^n$.
The syntax is
\begin{CodeSample}
GlobalResidueTheoremCPn[Num, Ideal, Vars]
\end{CodeSample}
where \Math{Num} denotes the numerator of $\omega$,
\Math{Ideal} the denominator factors and \Math{Vars} the variables.

As an example, let us consider the differential form in section~\ref{sec:GRT_example},
\begin{CodeSample}
Num=z[1];
Ideal={z[2] (a[1] z[1] + a[2] z[2]), (b[1] z[1] + b[2] z[2])};
Vars={z[1], z[2]};
GRTs=GlobalResidueTheoremCPn[Num, Ideal, Vars];
\end{CodeSample}
There are seven linear relations that arise from the global residue theorem,
\begin{CodeSample}
Length[GRTs]
Out: 7
\end{CodeSample}
The relations are recorded in the form $\{ P, \{ V, R \} \}$,
where $P$ denotes the denominator partition, $V$ the set of poles
involved in the relation and $R$ their respective residues.
For the denominator partition computed in detail in section~\ref{sec:GRT_example}
we have
\begin{CodeSample}
GRTs[[2]]
Out: {{w[0] w[2], (a[1] w[1] + a[2] w[2]) (b[1] w[1] + b[2] w[2])}, {{{1, 0, 0}, {0, 1, -(a[1]/a[2])}, {0, 1, -(b[1]/b[2])}}, {-(1/(a[1] b[1])), -(a[2]/(a[1] (-a[2] b[1] + a[1] b[2]))), -(b[2]/(b[1] (a[2] b[1] - a[1] b[2])))}}}
\end{CodeSample}
It is easy to check that the residues indeed sum to zero,
\begin{CodeSample}
Simplify[Total[GRTs[[2, -1, -1]]]]
Out: 0
\end{CodeSample}

\Math{GlobalResidueTheoremCPn} only records non-vanishing residues and
their respective poles. In cases where all the residues that appear
in a global residue theorem vanish, the empty set is returned as output.
\begin{CodeSample}
GlobalResidueTheoremCPn[1, {z[1] z[2] - 1, z[2]}, {z[1], z[2]}]
Out: {{{w[2], -w[0]^2 + w[1] w[2]}, {{}, {}}}}
\end{CodeSample}

\subsection{Options}
\label{sec:options}
The following options can be specified in \Math{MultivariateResidue}:

\vspace{2mm}\noindent$\bullet$
\Math{Method}: the type of algorithm used for computing multivariate residues.
Possible settings for this option are \Math{"TransformationFormula"} (default) and \Math{"QuotientRingDuality"}, described in sections \ref{sec:residues_from_transformation_formula} and \ref{sec:quotient_ring_duality}, respectively.

\vspace{2mm}\noindent$\bullet$
\Math{CoefficientDomain}: the type of objects assumed to be coefficients of monomials in the computation of Gr\"obner bases.
Possible settings for this option are \Math{InexactNumbers}, \Math{Rationals} and \Math{RationalFunctions} (default).

\vspace{2mm}\noindent$\bullet$
\Math{MonomialOrder}: the criterion used for monomial ordering in Gr\"obner basis computation and polynomial division. This option has an effect when using the method \Math{"TransformationFormula"} with \Math{$MultiResUseSingular=True}, or when using \Math{"QuotientRingDuality"}. Possible values are \Math{Lexicographic} (default), \Math{DegreeLexicographic} and \Math{DegreeReverseLexicographic}.

\vspace{2mm}\noindent$\bullet$
\Math{FindMinimumDelta}: compute and use the smallest possible $\delta$ entering the partition-of-unity polynomials $e_i=1-(1-L_i^\delta)^\delta$, which are needed to compute residues at all the finite poles of the variety. Possible settings are \Math{True} (default) and \Math{False}. Specifying \Math{True} will compute and use the smallest possible $\delta$; specifying \Math{False} will drop the calculation of a minimal $\delta$ and use a generally larger value (namely the maximum pole multiplicity).

\subsection{Global Settings}
\label{sec:globalsettings}
The following global settings can be specified in a Mathematica session that makes use of the package (the settings can be altered after loading the package and will impact all subsequent \Math{MultivariateResidue} evaluations):

\vspace{2mm}\noindent$\bullet$
\Math{$MultiResInputChecks}: check the input of \Math{MultivariateResidue}. It checks for instance whether the ideal is zero-dimensional. Possible values are \Math{True} (default) and \Math{False}. It can be switched off to improve efficiency.

\vspace{2mm}\noindent$\bullet$
\Math{$MultiResInternalChecks}: perform internal cross-checks, for instance on the correctness of the transformation matrix $A(z)$ and the partition of unity $e(i)$. Possible values are \Math{True} (default) and \Math{False}. Similarly, this can also be switched off to improve efficiency.

\vspace{2mm}\noindent$\bullet$
\Math{$MultiResUseSingular}: use Singular for computations of Gr{\"o}bner bases and polynomial division. Possible settings are \Math{True} and \Math{False} (default).
When set to \Math{True}, one must specify the path to the Singular executable in the variable \Math{$MultiResSingularPath}.
Singular is a computer algebra system dedicated to computational algebraic geometry that can outperform Mathematica for complicated calculations (see section \ref{sec:performance}) and therefore merits an interface between the two programs.
Singular is free software under the GNU General Public Licence. It may be obtained from \texttt{https://www.singular.uni-kl.de}.

\vspace{2mm}\noindent$\bullet$
\Math{$MultiResSingularPath}: the path to Singular. By default the path is set to \Math{"/usr/bin/Singular"}.

\subsection{Example of application: Cachazo-He-Yuan scattering equations}\label{sec:CHY_example}

As discussed in section~\ref{sec:applications}, multivariate residues
play a central role in the Cachazo-He-Yuan scattering equations which describe tree-level
scattering amplitudes in any spacetime dimension \cite{Cachazo:2013hca}.
In this formalism, an $n$-particle amplitude is expressed as
the $(n-3)$-fold contour integral in eq.~(\ref{eq:CHY_form_of_amplitude})
which localizes the integrand to the solution of the scattering equations
(\ref{eq:CHY_scattering_eqs}).

As observed in refs.~\cite{Sogaard:2015dba,Bosma:2016ttj}, the steps of
solving eq.~(\ref{eq:CHY_scattering_eqs}) and summing over the individual residues
can be circumvented by directly computing the global residue (\ref{eq:global_residue_def}).

\textsc{MultivariateResidues} allows direct computation of the global residue and thus
in turn tree amplitudes. To illustrate this, we consider the five-scalar tree amplitude
in $\phi^3$ theory. Following ref.~\cite{Sogaard:2015dba}, we enter the following input,\\[-6mm]
\begin{CodeSample}
 h[1] = s[12] + s[13]c[3] + s[14]c[4];
 h[2] = s[123]c[3] + s[124]c[4] + s[134]c[3]c[4];
gt[1] = (s[13]s[124] + s[13]s[134] - s[14] (s[123] + s[134]c[4]))/
         ((s[12] + s[13]) (s[124] + s[134]) - s[14]s[123]);
gt[2] = -((s[12]s[13]s[134] - (s[13] + s[14]) (s[13]s[124]
           -s[14] (s[123] + s[134]c[4])))/
         (s[12] (s[12]s[134] - (s[13] + s[14]) (s[123] + s[124]))));
gt[3] = (s[13] (s[124] + s[134]c[3]) - s[14]s[123])/(s[12]s[123]);

Vars   = {c[3], c[4]};
Num    = c[3] (1 - c[4]) gt[1] gt[2] gt[3];
Ideal = {h[1], h[2]};
\end{CodeSample}
where \texttt{(h[a], s[1a], s[1ab], c[a], gt[a], Num)} correspond respectively to
$(h_a, \sigma_a, \sigma_{ab}, z_a, \widetilde{g}_a, N(z_3,z_4))$ given
in eqs.~(3.5), (3.10)--(3.13) of ref.~\cite{Sogaard:2015dba}. In the above,
\texttt{s[1a]} and \texttt{s[1ab]} represent $(p_1 + p_a)^2$ and
$(p_1 + p_a + p_b)^2$. The functions \texttt{h[a]} represent
the $n - 3 = 2$ scattering equations (\ref{eq:CHY_scattering_eqs}) in polynomial form,
whereas \texttt{gt[a]} represent multiplicative inverses%
\footnote{The existence of these multiplicative inverses is guaranteed by Hilbert's Nullstellensatz.}
to the Parke-Taylor denominator factors in the integration measure.

The five-scalar amplitude can now be computed as a global residue,
\begin{CodeSample}
MultivariateResidue[Num, Ideal, Vars, {GlobalResidue},                                Method -> "QuotientRingDuality"]

Out:  (s[12]s[13]s[123]s[124] + s[12]s[14]s[123]s[124]                          + s[13]s[14]s[123]s[124] + s[14]^2 s[123]s[124]                                 - s[13]^2 s[124]^2 - s[13]s[14]s[124]^2 + s[12]s[13]s[123]s[134]        + s[12]s[13]s[124]s[134] - s[13]^2 s[124]s[134] - s[13]s[14]s[124]s[134]                 + s[12]s[13]s[134]^2)/                                                          (s[12]s[123] (-s[13]s[123] - s[14]s[123] - s[13]s[124]  - s[14]s[124]                         + s[12]s[134]) (-s[14]s[123] + s[12]s[124] + s[13]s[124]                         + s[12]s[134] + s[13]s[134]))
\end{CodeSample}
By applying momentum conservation identities, the output expression can be brought into the form,
\begin{equation}
A_5^{\mathrm{tree}, \hspace{0.6mm} \phi^3} = \frac{1}{s_{12} s_{34}} + \frac{1}{s_{23} s_{51}} + \frac{1}{s_{12} s_{45}}
               + \frac{1}{s_{34} s_{51}} + \frac{1}{s_{23} s_{45}} \,,
\end{equation}
which is manifestly the correct expression for the five-scalar tree amplitude
in $\phi^3$ theory.

\subsection{Performance}
\label{sec:performance}

The speed performance of \Math{MultivariateResidue} depends on the selected options and global settings; in particular the options \Math{Method}, \Math{MonomialOrder} and \Math{$MultiResUseSingular}.
This section demonstrates the impact of these options through a few explicit examples.

The default method \Math{"TransformationFormula"} is typically the best choice for simple problems, whereas the sophisticated \Math{"QuotientRingDuality"} can offer speed improvements in more involved computations.
For instance, the default method is the fastest method for the residue computation with the simple ideal defined in section \ref{sec:newfunction},
\begin{CodeSample}
First[AbsoluteTiming[MultivariateResidue[1, Ideal, Vars, Variety, Method -> "TransformationFormula"]]]
First[AbsoluteTiming[MultivariateResidue[1, Ideal, Vars, Variety, Method -> "QuotientRingDuality"]]]
Out: 0.255931
Out: 4.147544
\end{CodeSample}
On the other hand, if we consider an example with more variables, then the default method becomes less efficient (due to the costly computation of the transformation matrix), and one might opt for \Math{"QuotientRingDuality"},
\begin{CodeSample}
n = 10;
Ideal = Table[z[i], {i, 1, n}];
Vars = Table[z[i], {i, 1, n}]
Variety = {Table[0, {n}]};
First[AbsoluteTiming[MultivariateResidue[1, Ideal, Vars, Variety, Method -> "TransformationFormula"]]]
First[AbsoluteTiming[MultivariateResidue[1, Ideal, Vars, Variety, Method -> "QuotientRingDuality"]]]
Out: 1.029218
Out: 0.123257
\end{CodeSample}
Another circumstance under which \Math{"QuotientRingDuality"} performs better than \Math{"TransformationFormula"} is when points in the variety are large rational functions of one or more parameters. In such cases, the computation of Gr\"obner bases in \Math{"TransformationFormula"} can be rather slow. The following example with two complex variables illustrates this point.
\begin{CodeSample}
Ideal = {1 + c[1] z[1] + c[2] z[2], 1 + c[3] z[1] + c[4] z[2]};
Vars = {z[1], z[2]};
Variety = Map[Last, Sort[Solve[Ideal == 0, Vars]], {2}]
Out: {{-((c[2] - c[4])/(c[2] c[3] - c[1] c[4])),								  -((-c[1] + c[3])/(c[2] c[3] - c[1] c[4]))}}
\end{CodeSample}
For this problem, the \Math{"QuotientRingDuality"} method is faster,
\begin{CodeSample}
First[AbsoluteTiming[MultivariateResidue[1, Ideal, Vars, Variety, Method -> "TransformationFormula"]]]
First[AbsoluteTiming[MultivariateResidue[1, Ideal, Vars, Variety, Method -> "QuotientRingDuality"]]]
Out: 0.156636
Out: 0.074859
\end{CodeSample}

The method \Math{"QuotientRingDuality"} also allows the use of various monomial orderings in the computation of Gr{\"o}bner bases, which can impact the speed of subsequent polynomial reductions.
Considering once again the ideal defined in section \ref{sec:newfunction} and computing the multivariate residue using three different monomial orderings,
\begin{CodeSample}
Ideal = {z[1]^2 (1 + z[1] - z[2]), z[3]^3,
   		z[2]^3 (-1 - z[1] - a z[1] + z[2] + a z[1] z[3])};
Vars = {z[1], z[2], z[3]};
Variety = Map[Last, Sort[Solve[Ideal == 0, Vars]], {2}];
First[AbsoluteTiming[MultivariateResidue[1, Ideal, Vars, Variety, Method -> "QuotientRingDuality", MonomialOrder -> #]]] & /@ {Lexicographic, DegreeLexicographic, DegreeReverseLexicographic}
Out: {4.128913, 1.459281, 1.733149}
\end{CodeSample}
shows that the options \Math{DegreeLexicographic} and \Math{DegreeReverseLexicographic} are noticeably faster than \Math{Lexicographic}.
The reason for the speed difference (which becomes more pronounced upon raising the powers of factors in the ideal) is that the \Math{Lexicographic} monomial ordering produces a Gr{\"o}bner basis for the ideal which contains only four polynomials (with $2, 3, 4$ and $16$ terms), whereas the other two monomial orderings produce a Gr{\"o}bner basis for this ideal with eight polynomials (with $2, 2, 3, 3, 4, 5, 16$ and $19$ terms).
The polynomial reduction is faster when using the larger Gr{\"o}bner basis.

\begin{figure}[!h]
\includegraphics[width=\textwidth]{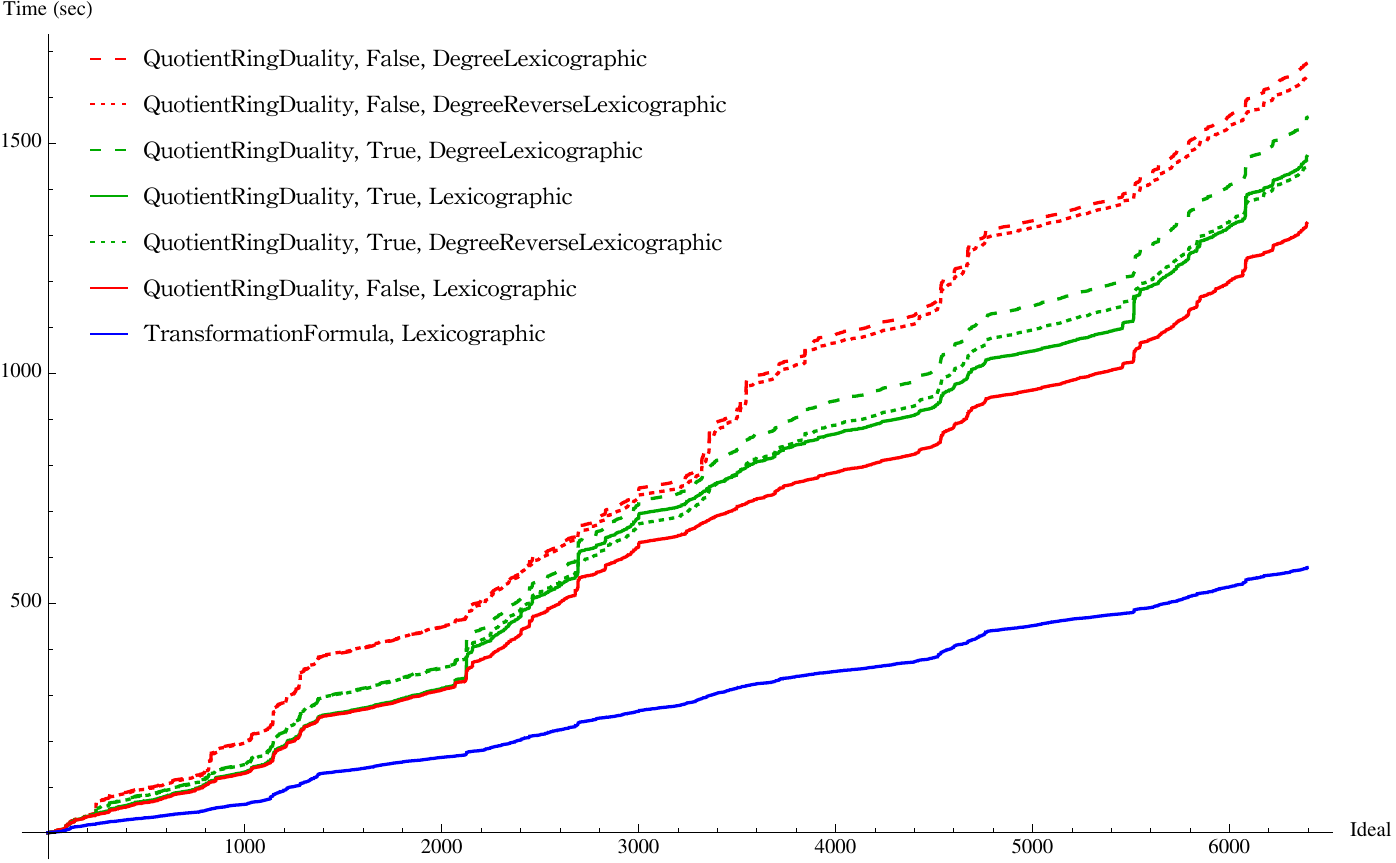}
\caption{
(Color online.) The cumulative evaluation time of residues (vertical axis) over the set of 6395 ideals (horizontal axis) associated with the pentacut of a generic gauge theory two-loop four-gluon integrand for the different residue computation options in the package \textsc{MultivariateResidues}.
The legend items are ordered by descending total computation time.
The plateaus present in \Math{"QuotientRingDuality"} show that the underlying computational procedure is sensitive to the precise structure of the ideal.
In contrast, \Math{"TransformationFormula"} exhibits a more linear profile, showing that the underlying computational procedure is less sensitive to the precise structure of the ideal.
We also performed these computations with \Math{$MultiResUseSingular=True}, which increased performance by 18\% on average.
(Color online. These timings were obtained on a single laptop with 8 GB RAM and 1.8 GHz Intel Core i5 processor.)}
\label{fig:CutTwoLoopIntegrandEvaluationTimes}
\end{figure}

Figure~\ref{fig:CutTwoLoopIntegrandEvaluationTimes} shows the cumulative evaluation time of residues with the various options of \Math{MultivariateResidue} for the case of a four-gluon two-loop integrand arising after the pentacut shown in figure~\ref{fig:cut_slashed_box} has been applied.
In such problems, where one faces many residue computations, a straightforward way to decrease the total running time is by performing the computations in parallel. This can be achieved by including
\begin{CodeSample}
DistributeDefinitions["MultivariateResidues`"];
\end{CodeSample}
and making use of the function \Math{ParallelMap}.
(No parallelization was used in producing figure~\ref{fig:CutTwoLoopIntegrandEvaluationTimes}.)

Finally, it is worth mentioning that multivariate residue calculations can be simplified -- where possible -- by appropriate use of partial fractioning.
Suppose one computes the residues of $R(z_1,z_2) = (z_1 f_1+ z_2 f_3)/(f_1\,f_2\,f_3\,f_4)$, where $f_i = a_i z_1 + b_i z_2$, with respect to the ideal $I(z_1,z_2) = \langle \, f_1 f_2 \,, f_3 f_4\, \rangle$.
A direct computation of the residue yields
\begin{CodeSample}
f[i_] := a[i] z[1] + b[i] z[2];
Res[1] = MultivariateResidue[z[1] f[1] + z[2] f[3], {f[1] f[2], f[3] f[4]}, {z[1] -> 0, z[2] -> 0}]
Out: -((a[3] a[4] b[2] + a[4] b[1] b[2] - a[2] a[4] b[3] - a[1] b[2] b[4])/((a[3] b[2] - a[2] b[3]) (a[4] b[1] - a[1] b[4]) (a[4] b[2] - a[2] b[4])))
\end{CodeSample}
Partial fractioning the rational function produces two terms, $R(z_1,z_2) = z_1/(f_2\,f_3\,f_4) + z_2/(f_1\,f_2\,f_4)$. The residue of each of these terms should be computed with respect to the corresponding reduced ideals $\langle \, f_2 \,, f_3 f_4 \, \rangle$ and $\langle \, f_1 f_2 \,, f_4 \, \rangle$.
The sum of these two residues reproduces the result of the direct computation,

\begin{CodeSample}
Res[2, 1] = MultivariateResidue[z[1], {f[2], f[3] f[4]}, {z[1] -> 0, z[2] -> 0}]
Res[2, 2] = MultivariateResidue[z[2], {f[1] f[2], f[4]}, {z[1] -> 0, z[2] -> 0}]
Simplify[Res[1] == Res[2, 1] + Res[2, 2]]
Out: -(b[2]/((a[3] b[2] - a[2] b[3]) (a[4] b[2] - a[2] b[4])))
Out: -(a[4]/((a[4] b[1] - a[1] b[4]) (a[4] b[2] - a[2] b[4])))
Out: True
\end{CodeSample}
and is faster than the direct computation (in this case 0.0882 seconds versus 0.4014 seconds).

\vspace{2mm}

\section{Conclusions}\label{sec:Conclusions}

In this paper we have introduced the Mathematica package
\textsc{MultivariateResidues} for the evaluation of multivariate residues.
The implementation can be used to compute any multivariate
residue of any rational form, including the case where the denominator ideal involves
parameters (corresponding, for example, to Lorentz invariants of external
momenta in a scattering process).

We have implemented two different algorithms for the computation of residues,
one (\Math{"TransformationFormula"}) based on the transformation formula
and one (\Math{"QuotientRingDuality"}) exploiting that the residue map
defines a non-degenerate inner product on the quotient ring.

We have applied our code to 6500 examples arising in the computation of
generalized-unitarity cuts of two-loop scattering amplitudes. From doing so,
we have observed the following patterns regarding the relative performance of the
two algorithms. In cases where the ideals involve only few parameters, and the
expressions of the parameters are polynomials of low degree, \Math{"TransformationFormula"}
is the faster option. Moreover, when the
cumulative computation time of \Math{"TransformationFormula"} is plotted against
the range of ideals over which it has been applied, it displays a more linear profile
than \Math{"QuotientRingDuality"}, showing that the underlying computational
procedure is rather insensitive to the precise structure of the ideal.

In contrast, \Math{"QuotientRingDuality"} tends to perform better in more
involved computations because this circumvents the need to compute the transformation matrix
$A(z)$ in eq.~\eqref{eq:transformation_law}. This becomes increasingly apparent in
cases with several variables, or with several parameters, or
whenever ideals involve high-degree polynomials in the parameters.

\section*{Acknowledgments}
We thank Yang Zhang for useful discussions and Romain M{\"u}ller for assistance
with drawing figure \ref{fig:non-homologous_contours}.
The research leading to these results has received funding from the European Union Seventh Framework Programme (FP7/2007-2013) under grant agreement no. 627521, and from the Foundation for Fundamental Research of Matter (FOM), programme 156, ``Higgs as Probe and Portal'' and the Dutch National Organization for Scientific Research (NWO).
The work of KJL is also supported by ERC-2014-CoG, Grant number 648630 IQFT.

\appendix

\section{Topology of multivariate residues}\label{sec:topology_of_multivariate_residues}

In this appendix we aim to explain the underlying topological reason
why the value of a multivariate residue is not uniquely determined
by the pole enclosed by the integration cycle, but also depends on the cycle.
To gain a concrete understanding, we will consider the example of
the differential form $\omega$ in eq.~\eqref{eq:degenerate_diff_form}.
The presentation here is largely based on that of section II B of
ref.~\cite{Johansson:2015ava} by one of the present authors.

As computed in section~\ref{sec:residues_from_transformation_formula},
the differential form \eqref{eq:degenerate_diff_form} has the three distinct residues at
$p=(0,0)$ given in eqs.~\eqref{eq:R_1_def}--\eqref{eq:R_3_def}. As we will
see shortly, this is reflected in the fact that there are several
distinct integration cycles based at $p$ which yield distinct residues.
The higher-dimensional situation is thus quite different from contour integration
in one complex variable, where a contour either encloses a pole or doesn't,
and there is a unique value for the residue.

To clarify the situation, we seek to split eq.~\eqref{eq:degenerate_diff_form}
into terms with two distinct denominator factors. To this end, we
make the following change of variables,
\begin{equation}
\begin{aligned}
z_1' &= a_1 z_1 + a_2 z_2 \\
z_2' &= z_2 \,.
\end{aligned}
\end{equation}
After applying this transformation and dropping the primes
on $z_i$, the form then becomes,
\begin{align}
\omega = \frac{1}{a_1}
\left(\frac{1}{z_2 (c_1 z_1 + c_2 z_2)} - \frac{a_2}{z_1 (c_1 z_1 + c_2 z_2)}\right)
\d z_1\wedge \d z_2\,,
\label{eq:omega_partial_fractioned}
\end{align}
where $c_1 \equiv b_1$ and $c_2 \equiv a_1 b_2 - a_2 b_1$.
(This separation is a partial fractioning  followed by a change of variables.)

We start by examining the first term of eq.~\eqref{eq:omega_partial_fractioned}.
The canonical integration contour is a product of two circles. Choosing each circle
to be based at each denominator factor,
\begin{align}
|z_2|               &=  \delta_2 \\
|c_1 z_1 + c_2 z_2| &=  \delta_c \,,
\end{align}
where $\delta_2, \delta_c > 0$, the residue of this term is,
\begin{equation}
\frac{1}{a_1 b_1}\,,
\end{equation}
independent of the precise values of the radii of the circles.
To be a bit more explicit, we can parametrize the integration cycle as
\begin{equation}
\sigma{:} \hspace{6mm} \left\{ \hspace{-0.7mm}
\begin{array}{rl}
z_2                \hspace{1mm} =&  \hspace{-1mm} \delta_2 e^{i\theta_2} \\[1.5mm]
c_1 z_1 + c_2 z_2  \hspace{1mm} =&  \hspace{-1mm} \delta_c e^{i\theta_c} \,,
\label{eq:parametrization_of_torus}
\end{array}
\right.
\end{equation}
so that, as $\theta_2$ and $\theta_c$ run over the interval $[0,2\pi]$, the cycle is
traced out. By parametrizing $\int_\sigma \omega$ in this way, the integral can then
be evaluated as an ordinary two-fold integral over $\theta_2$ and $\theta_c$.

We now turn to the second term of eq.~\eqref{eq:omega_partial_fractioned}.
We must now examine whether any choices of the radii $\delta_2$ and $\delta_c$
leave the integrand singular on $\sigma$, as these would define an
\emph{illegitimate} integration cycle. The second denominator factor is of course
nonvanishing on the cycle~(\ref{eq:parametrization_of_torus}).
For the first factor we obtain from the parametrization (\ref{eq:parametrization_of_torus}),
\begin{align}
z_1 = \frac{1}{c_1} \big( \delta_c e^{i\theta_c} - c_2 \delta_2 e^{i\theta_2} \big) \,.
\label{eq:z_1_contour}
\end{align}
We observe that the first denominator factor will not vanish as long
as $\delta_c\neq |c_2|\delta_2$. On the other hand, if
$\delta_c = |c_2|\delta_2$, $z_1$ is guaranteed to vanish for some values of
the angles. The illegitimate choice $\delta_c = |c_2|\delta_2$ therefore divides the
moduli space $(\delta_c,\delta_2)$ into two regions,
\begin{equation}
\begin{aligned}
\mathrm{(1){:}} \hspace{4mm} \delta_c &> |c_2|\delta_2 \\[1mm]
\mathrm{(2){:}} \hspace{4mm} \delta_c &< |c_2|\delta_2\,,
\label{eq:regions_of_torus_moduli_space}
\end{aligned}
\end{equation}
which we will consider in turn.

We start by observing from the second equation of eq.~\eqref{eq:parametrization_of_torus}
that the parameter $\theta_c$ traces out a circle around the zero of the second denominator
factor. For the cycle $\sigma$ to enclose the pole at $(z_1, c_1 z_1 + c_2 z_2) = (0,0)$,
the issue is therefore whether the zero of the first denominator factor is encircled
by the \emph{other} independent parameter $\theta_2$. That is, whether for a fixed value of $\theta_c$,
the contour~\eqref{eq:z_1_contour} traced out by $\theta_2$ encloses $z_1 = 0$ or not.

Now, in region~(1), for a fixed value of $\theta_c$, the contour~\eqref{eq:z_1_contour}
traced out by $\theta_2$ is a circle centered at $\delta_c e^{i\theta_c}$.
As the radius $|c_2| \delta_2$ is less than $\delta_c$, the circle fails to enclose
$z_1 =0$. For the torus $\sigma$, this translates into saying that the pole $(z_1, z_2) = (0,0)$ is sitting
at the center of the symmetry plane of $\sigma$, but not inside the ``tube''.
We conclude that in region~(1), the second term in eq.~\eqref{eq:omega_partial_fractioned}
integrated over the cycle \eqref{eq:parametrization_of_torus} produces a vanishing residue.

In contrast, in region~(2), the $\theta_2$-parametrized contour does encircle
$z_1 =0$, and hence the second term in eq.~\eqref{eq:omega_partial_fractioned}
integrated over $\sigma$ produces a nonvanishing residue.
In particular, we observe that the residue of eq.~\eqref{eq:omega_partial_fractioned}
differs in the two regions \eqref{eq:regions_of_torus_moduli_space} and thus depends on
the relative radii $\delta_2$ and $\delta_c$ of the integration cycle.

More generally, let us consider a generic torus,
\begin{equation}
\begin{aligned}
z_1 &= \delta_{1,1} e^{i\theta_1} + \delta_{1,2} e^{i\theta_2} \\
z_2 &= \delta_{2,1} e^{i\theta_1} + \delta_{2,2} e^{i\theta_2} \,,
\label{eq:generic_torus}
\end{aligned}
\end{equation}
where the $\delta_{i,j}$ are real positive constants which determine the shape of the
cycle. For the two-form at hand, we can rescale all the $\delta_{i,j}$ uniformly
without loss of generality, so that we only have three independent real
parameters.

The integration cycle is legitimate for the first term in eq.~\eqref{eq:omega_partial_fractioned}
if and only if $\delta_{2,1} \neq \delta_{2,2}$ and $r_1 \neq r_2$, where
\begin{align}
r_1 = |c_1\delta_{1,1}+c_2\delta_{2,1}| \hspace{6mm} \mathrm{and} \hspace{6mm}
r_2 = |c_1\delta_{1,2}+c_2\delta_{2,2}|\,.
\end{align}
The cycle is legitimate for the second term if and only if
$\delta_{1,1}\neq\delta_{1,2}$ and $r_1 \neq r_2$. Thus, we must consider
eight regions, corresponding to choosing the upper or lower
inequality in each of the three relations,
\begin{equation}
\delta_{2,1}\gtrless\delta_{2,2}\,, \hspace{5mm}
\delta_{1,1}\gtrless\delta_{1,2}\,, \hspace{5mm}
r_1\gtrless r_2\,.
\label{eq:moduli_space_boundaries}
\end{equation}
We denote the upper choice by `$+$' and the lower choice by `$-$'. Each
region is then labeled by a string of signs. We see that in the region $M^{+++}$,
corresponding to $\delta_{2,1} > \delta_{2,2}$, $\delta_{1,1} > \delta_{1,2}$ and $r_1 > r_2$,
the zeros of all denominator factors of the two terms in eq.~\eqref{eq:omega_partial_fractioned}
are all encircled by the parameter $\theta_1$, so that the torus fails to enclose
the pole of either term and hence produces a vanishing residue.
In $M^{++-}$, the torus will enclose both terms, and the residue will be the sum of
the two terms' residues. In $M^{+-+}$, the torus only encloses the second term,
and in $M^{+--}$, the torus only encloses the first term. The remaining four regions
are related to these four by flipping all inequalities which leaves the results invariant (up to a sign).

The above analysis shows that multivariate residues are in general not fully
characterized by the location of the pole. Rather, the value of the residue depends
also on the shape of the cycle enclosing the pole. In the present example we
found that the moduli space of allowed integration cycles is divided into
several regions. These regions correspond to distinct homology classes of the
$(z_1, z_2)$ space
\begin{equation}
\mathbb{C}^2 \setminus \bigcup_{i=1}^3 D_i \,,
\end{equation}
where each $D_i \equiv \{ (z_1, z_2) \in \mathbb{C}^2 : \varphi_i (z_1, z_2)=0 \}$ is the surface
where the $i$th denominator factor of $\omega$ vanishes
(cf.~eq.~\eqref{eq:denominator_factors_example}) and $\omega$ hence is not well-defined.
The surfaces $D_i$ are called the \emph{divisors} of $\omega$.

\begin{figure}[!htb]
    \centering
    \begin{minipage}{.5\textwidth}
        \flushleft
        \includegraphics[width=1.2\linewidth, height=0.25\textheight]{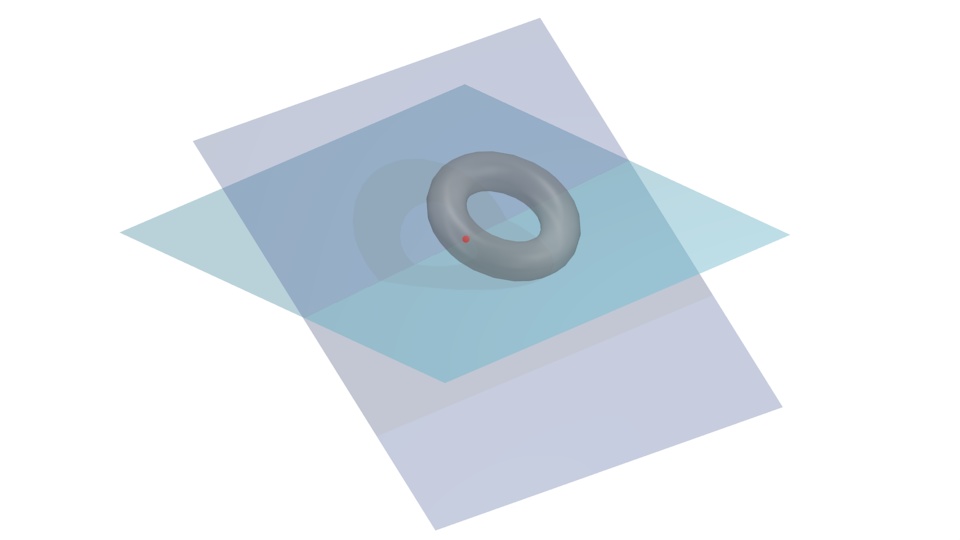}
    \end{minipage}%
    \begin{minipage}{0.5\textwidth}
        \includegraphics[width=1.2\linewidth, height=0.25\textheight]{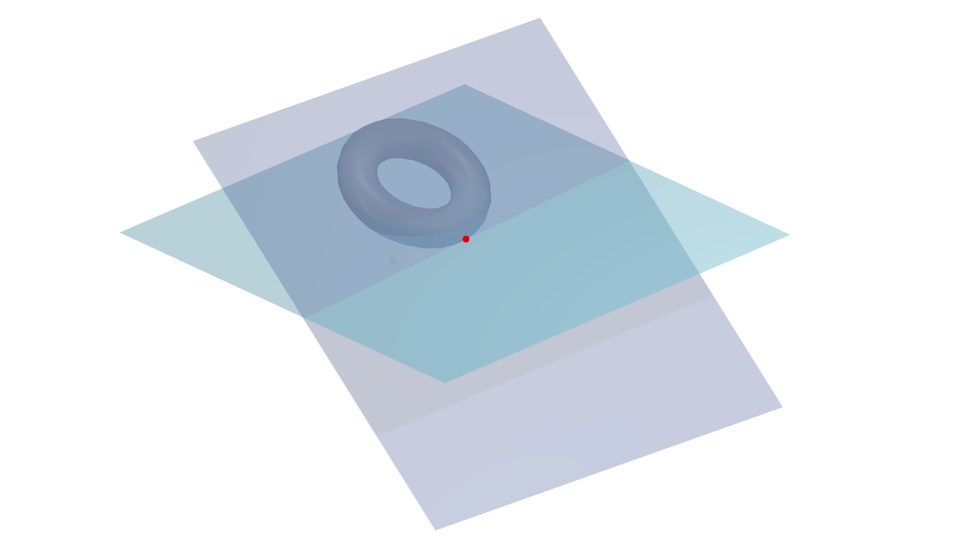}
    \end{minipage}
\vspace{-2mm}
\caption{(Color online.) A schematic illustration of the topological reason
why the value of a multivariate residue is not uniquely determined
by the pole enclosed by the integration cycle, but also depends on the cycle.
The figure is a projection of the two-complex-dimensional (i.e., four-real-dimensional)
situation for the differential form $\omega$ in eq.~\eqref{eq:degenerate_diff_form}.
The planes represent the divisors of $\omega$.
For visual clarity only two planes are shown. The red dot represents the pole
$p = (0,0)$, and the tori two distinct integration cycles enclosing the pole.
Because of the presence of the divisors, the cycle in the left figure cannot
be continuously deformed into the one in the right figure: the cycles are non-homologous.
As a result, the contours are not guaranteed to produce equal residues, and in general they will not.
It is an artifact of the two-dimensional projection of the four-dimensional situation
that the torus in the right figure appears to intersect the divisors and not to enclose
the pole.}
\label{fig:non-homologous_contours}
\end{figure}

Integration cycles with moduli taken from distinct regions $M^{+++}, M^{++-}$ etc.~are
\emph{non-homologous}, and as a result are not guaranteed to produce identical residues.
Figure~\ref{fig:non-homologous_contours} gives a schematic representation of the
divisors of $\omega$ and two non-homologous integration contours.

We can apply the residue evaluation algorithm of section~\ref{sec:residues_from_transformation_formula}
to each of the two terms in eq.~(\ref{eq:omega_partial_fractioned}) separately,
yielding $R_1$ and $R_2$ for the first and second term, respectively.
(Recall eqs.~\eqref{eq:R_1_def}--\eqref{eq:R_2_def} for the expressions for $R_1$ and $R_2$.)
Combining this with the observations
made in the discussion below eq.~\eqref{eq:moduli_space_boundaries}, we see
that in the region $M^{+++}$ of the moduli space, the residue evaluates to
$0$; in $M^{+--}$ to $R_1$; in $M^{+-+}$ to $R_2$; and in $M^{++-}$ to
$R_1 + R_2 = -R_3$ (cf.~eq.~\eqref{eq:residue_relation}).
From these observations we conclude that we have the following one-to-one map between the partitionings
of the denominator of $\omega$ in eq.~\eqref{eq:denominator_partitionings_example}
and the regions of the moduli space of integration cycles,
\begin{equation}
\begin{aligned}
\{ \varphi_1, \varphi_2 \varphi_3 \}  \hspace{2mm}&\longleftrightarrow\hspace{2mm}  M^{+--} \\
\{ \varphi_2, \varphi_3 \varphi_1 \}  \hspace{2mm}&\longleftrightarrow\hspace{2mm}  M^{+-+} \\
\{ \varphi_3, \varphi_1 \varphi_2 \}  \hspace{2mm}&\longleftrightarrow\hspace{2mm}  M^{++-} \,.
\end{aligned}
\end{equation}
This map provides a dictionary between the algebraic and geometric
pictures of the distinct residues defined at the given pole.

We remark that the relation \eqref{eq:residue_relation} shows that
only two of the regions $M^{+--}, M^{+-+},\ldots$ define
linearly independent integration cycles.

\bibliographystyle{elsarticle-num}




\end{document}